\documentclass[aps,english,preprint,nofootinbib,preprintnumbers]{revtex4}
\usepackage{mathrsfs}
\usepackage{hyperref}
\usepackage{graphicx}
\usepackage{amsmath, amssymb}
\usepackage{babel}
\usepackage{color}
\usepackage{slashed}

\def\beqn{\begin{eqnarray}}
\def\eeqn{\end{eqnarray}}
\def\barr{\begin{array}}
\def\earr{\end{array}}
\def\btab{\begin{tabular}}
\def\etab{\end{tabular}}
\def\bite{\begin{itemize}}
\def\eite{\end{itemize}}
\def\bcen{\begin{center}}
\def\ecen{\end{center}}

\begin{document}

\title{Electron Mass Singularities in Semileptonic Kaon Decays}

\author{Chien-Yeah Seng$^{1}$}
\author{William J. Marciano$^{2}$}
\author{Ulf-G. Mei{\ss}ner$^{1,3,4}$}

\affiliation{$^{1}$Helmholtz-Institut f\"{u}r Strahlen- und Kernphysik and Bethe Center for
	Theoretical Physics,\\
	Universit\"{a}t Bonn, 53115 Bonn, Germany}
\affiliation{$^{2}$Department of Physics, Brookhaven National Laboratory, Upton, New York 11973, USA}
\affiliation{$^{3}$Institute for Advanced Simulation, Institut f\"ur Kernphysik and J\"ulich
	Center for Hadron Physics,
	Forschungszentrum J\"ulich, 52425 J\"ulich, Germany}
\affiliation{$^{4}$Tbilisi State  University,  0186 Tbilisi, Georgia}

\date{\today}

\begin{abstract}

We show that recent improvements in the $O(\alpha)$ long-distance quantum electrodynamics (QED) corrections to the radiative inclusive $K_{e3}$ decay rate using the Sirlin representation are free from infrared divergences and collinear electron mass singularities in the limit $m_e\rightarrow0$, as predicted by the Kinoshita-Lee-Nauenberg theorem.  We also verify that in massless QED with the simultaneous dimensional regularization of QED photon infrared divergences and electron mass singularities
leads to the same result for the inclusive rate in the limit of four space-time dimensions. The equivalence of the two approaches results in part from an interesting interplay between a small chirality-breaking effect in the massless electron limit and the generalization of space-time algebra and phase space integrals to $d>4$ dimensions. Our finding supports the small theoretical uncertainty claimed for $K_{e3}$ radiative inclusive rates and reaffirms its utility in precision unitarity tests of the quark mixing matrix.
\end{abstract}

\maketitle


\section{Introduction}

Radiative inclusive semileptonic kaon decays into electrons or muons along with soft or hard bremsstrahlung ($K\rightarrow \pi\ell\nu(\gamma)$, denoted as $K_{\ell 3}$) provide one of the most precise determinations of the first-row Cabibbo-Kobayashi-Maskawa~\cite{Cabibbo:1963yz,Kobayashi:1973fv} quark mixing matrix element $|V_{us}|$.  The existence of 6 distinct charged and neutral such $K$ 
decay modes allows consistency checks and averaging that currently result 
in a $\pm 0.2\%$ determination of $|V_{us}|$.
That quantity can be identified with sine of the Cabibbo angle in
the effective (two generations) four quark limit. It plays an   
essential role in precision unitarity tests of the Standard Model (SM).  

In recent years, a significant roughly 1\% difference has been observed between the values of $|V_{us}|$ obtained from $K_{\ell 3}$ in comparison with the radiative inclusive leptonic kaon decay mode ($K\rightarrow \mu\nu(\gamma)$, denoted as $K_{\mu 2}$)~\cite{Seng:2022wcw}:
\begin{equation}
|V_{us}|=\left\{ \begin{array}{ccc}
0.22308(55) & , & K_{\ell3}\\
0.22520(50) & , & K_{\mu2}
\end{array}\right.~.
\end{equation} 
The discrepancy is about 2.9$\sigma$. Similarly, improvements in the calculation of the inner radiative corrections (RC) to free neutron and nuclear beta decays~\cite{Seng:2018yzq,Seng:2018qru,Czarnecki:2019mwq,Gorchtein:2018fxl,Seng:2020wjq,Shiells:2020fqp,Hayen:2020cxh} reduce the central value of $|V_{ud}|$~\cite{Hardy:2020qwl} and its theoretical uncertainty, resulting in an apparent violation of the first-row CKM unitarity relation $|V_{ud}|^2+|V_{us}|^2+|V_{ub}|^2=1$ (with negligible $|V_{ub}|^2$) up to about 3.2$\sigma$~\cite{Seng:2021nar}. These various inconsistencies between different precision experimental  extractions of the Cabibbo angle $\theta_C=\sin^{-1}|V_{us}|$ and the apparent violation of first row CKM unitarity are commonly referred to as the ``Cabibbo angle anomaly''. Its possible implications for ``new physics'' beyond the Standard Model (BSM) expectations have been extensively discussed in the  literature~\cite{Gonzalez-Alonso:2018omy,Bryman:2019ssi,Bryman:2019bjg,Belfatto:2019swo,Tan:2019yqp,Grossman:2019bzp,Coutinho:2019aiy,Falkowski:2019xoe,Cirgiliano:2019nyn,Jho:2020jsa,Yue:2020wkj,Endo:2020tkb,Capdevila:2020rrl,Eberhardt:2020dat,Cheung:2020vqm,Crivellin:2020ebi,Crivellin:2020klg,Crivellin:2020lzu,Kirk:2020wdk,Crivellin:2020oup,Falkowski:2020pma,Becirevic:2020rzi,Crivellin:2021njn}.

To confirm or negate the current disagreements between experiments and SM theory predictions, highly precise theory inputs are needed. In $K_{\ell 3}$ decays, they include the    $K^0\pi^-$ transition form factor $f_+^{K^0\pi^-}(0)$ at zero momentum transfer~\cite{Lubicz:2009ht,Bazavov:2012cd,Boyle:2015hfa,Carrasco:2016kpy,Bazavov:2018kjg} (determined to about 0.2\% using lattice QCD and currently representing the dominant theory uncertainty), the phase space factors~\cite{Lichard:1997ya,Antonelli:2010yf,Hill:2006bq,Bernard:2006gy,Bernard:2009zm,Abouzaid:2009ry,PSCKM21}, quark mass parameters used to estimate the isospin-breaking correction for
 $f_+^{K^+\pi^0}$ (mainly due to $\pi^0\eta$-mixing)  ~\cite{RBC:2014ntl,Durr:2010vn,Durr:2010aw,MILC:2009ltw,Fodor:2016bgu,Bazavov:2017lyh,EuropeanTwistedMass:2014osg,FermilabLattice:2014tsy,Giusti:2017dmp,Colangelo:2018jxw}, and the long-distance electromagnetic radiative corrections (EMRC). In particular, a recent re-analysis of the $K_{\ell 3}$ EMRC~\cite{Seng:2021boy,Seng:2021wcf,Seng:2022wcw} based on a new theory framework~\cite{Seng:2019lxf,Seng:2020jtz} that hybridizes the classic Sirlin representation~\cite{Sirlin:1977sv,Seng:2021syx}, the modern language of Chiral Perturbation Theory (ChPT) and the latest lattice calculations of mesonic $\gamma W$-box diagrams~\cite{Feng:2020zdc,Ma:2021azh} achieves a precision level of $O(10^{-4})$, an order of magnitude better than the existing pure ChPT analyses~\cite{Cirigliano:2001mk,Cirigliano:2004pv,Cirigliano:2008wn}. These new results sharpen the $K_{\ell3}$--$K_{\mu 2}$ discrepancy in the $|V_{us}|$ extractions.

At the $O(10^{-4})$ level, every approximation and uncertainty estimate made in the theory analysis must be thoroughly scrutinized. In particular, the procedure of separating the EMRC into perturbative and non-perturbative pieces must be compatible with general properties of quantum field theory.  Measurable (infrared safe) quantities must be properly regularized such that QED photonic ``infrared (IR)-divergences'' due to (virtual and real) emission of low-energy photons cancel. For decays such as $K_{e3}$ with a light electron in the final state, ``mass singularities'' (i.e. terms enhanced by $\ln m_e$) due to the emission of a photon collinear to the positron (or electron) are generally present in the differential decay rate. However, such terms which are divergent in the massless
limit ($m_e=0$) cancel in the total integrated radiative inclusive decay rate. Since the experimental values of the $K_{e3}$ partial widths are derived from kaon lifetime measurements which are by definition radiative inclusive and experimental branching ratios, which are assumed to be fully radiative inclusive, theory calculations used for comparison must, therefore, also be radiative inclusive.

The cancellation of infrared photonic divergences between real and virtual radiative corrections was realized as early as the 1930s by Bloch and Nordsieck~\cite{Bloch:1937pw}, and later generalized  by Yennie, Frautschi and Suura~\cite{Yennie:1961ad}. The cancellation of electron mass singularities in radiative inclusive decays was first observed by Kinoshita and Sirlin in their calculation of QED corrections to the muon total decay rate in the Fermi V-A theory~\cite{Kinoshita:1958ru}. Later generalization of that feature is known as the Kinoshita-Lee-Nauenberg (KLN) theorem~\cite{Kinoshita:1962ur,Lee:1964is}. The theorem is valid to all orders in perturbation theory for fully inclusive rate  expansions in terms of bare couplings and masses.  However, renormalization can induce mass singularities. For example, electric charge renormalization at zero momentum transfer used to define $\alpha$, the fine structure constant, if used as the expansion parameter will induce two loop                           
$\ln m_e$ dependent corrections.  Based on that connection, Roos and Sirlin derived the leading $\ln m_e$ logarithmic QED radiative corrections to the muon lifetime that start to appear at two loop order ~\cite{Roos:1971mj,Marciano:1975de}.

In Refs.~\cite{Seng:2021boy,Seng:2021wcf,Seng:2022wcw}, the cancellation of the photonic IR divergences is explicitly demonstrated for $K_{e3}$ by computing the so-called ``convection term'' contribution~\cite{Yennie:1961ad,Meister:1963zz} to the virtual and real corrections analytically. In Sirlin's representation, that term represents the complete IR divergence, but not the full electron mass singularities contribution. The missing part was computed numerically using the physical electron mass as input. Therefore, the expected cancellation of electron mass singularities in the total numerical result for the inclusive decay rate was not directly tested. In principal, some $\ln m_e$-enhanced contributions could have been missed or lost in the approximations made.  If so, it could invalidate the final error analysis and might shift the theoretical result somewhat.  Indeed, the explicit cancellation of mass singularities in radiative inclusive processes is a useful tool for checking difficult calculations.   

The first part of this paper serves to fill in this missing check. By extending the analytically-calculable terms in both virtual and real corrections and expanding them with respect to $m_e$ (which we will call the ``mass-expansion'' method), we demonstrate explicitly that the approximations made in Refs.~\cite{Seng:2021boy,Seng:2021wcf,Seng:2022wcw} within Sirlin's representation indeed result in a singularity-free total $K_{e3}$ decay rate. A similar study was done previously in Ref.~\cite{Bytev:2002nx} but only in the $K_{e3}^+$ channel.  Our analysis constitutes a more general proof. 
Since cancellation is achieved in the total decay rate, this calculation provides a nice check not just to the validity of the approximations made on the interaction dynamics at the level of the squared amplitude, but also the correctness of the kinematic settings of the three- and four-body phase space integrations. 

The second part of this paper is more theoretically oriented. It examines the possibility of employing $M_\gamma=m_e=0$, i.e. massless QED, from the start of a radiative inclusive calculation. Because the kinematics are simpler than the massive electron calculation, it represents a simplified method of checking the  $K_{e3}$ decay rate in the limit of zero electron mass. 

For such studies, a regulator that preserves the properties of QED must be used. A powerful technique for this purpose is the continuous dimensional regularization (DR) approach~\cite{tHooft:1972tcz,Bollini:1972ui,Ashmore:1972uj,Gastmans:1973uv,Marciano:1974tv,Marques:1974ab,Marciano:1974kw,Marciano:1975de} which generalizes the space-time dimension to $d=4-\epsilon$. IR and mass singularities manifest themselves as poles of order $(2/\epsilon)^2$ and $(2/\epsilon)$. Note that $\epsilon$ is negative for ${d>4}$, the space-time domain where infrared effects are finite. This method is used extensively in the study of perturbative Quantum Chromodynamics (pQCD) with massless quarks (see, e.g., \cite{Collins:2011zzd,Becher:2014oda} and references therein). An interesting comparison is the relationship between the prediction of singularity-free quantities for the $\epsilon\rightarrow 0$ limit in the DR method and the $m_e\rightarrow 0$ limit in the mass-expansion method. Ref.~\cite{Marciano:1975de} demonstrated the equivalence between these two methods in the total radiatively inclusive $W\rightarrow e\nu(\gamma)$ decay rate, but similar comparisons in processes with more complicated kinematics like $K_{e3}$ have, to the best of our knowledge, not been studied before. They are of interest both in terms of testing the robustness of the regularization techniques and achieving a better understanding of the general properties of the underlying quantum field theory. 

In this paper we demonstrate that the mass-expansion and DR methods both satisfy the KLN theorem and give the same result for the total radiatively inclusive $K_{e3}$ decay rate in the massless electron limit. That finding is highly non-trivial, particularly for the final finite decay rate prediction. Two essential steps are needed: In the case of DR, a generalization of the three- and four-body phase space integrals along with algebraic manipulations lead to an ${\epsilon}$ dependence which produces a finite contribution in the limit ${\epsilon=0}$.
In the mass expansion approach, terms in the bremsstrahlung process that are proportional to $(p_e\cdot k)^{-2}$, with $k$ the photon momentum must be kept even though they are suppressed by $m_e^2$ in the squared amplitude. They  acquire a $1/m_e^2$-enhancement upon performing the phase space integration which lifts the suppression and results in a small  finite contribution in the $m_e\rightarrow 0$ limit. When all such effects are properly taken into account, the finite parts in the two methods are in perfect agreement. 

The contents of this paper are arranged as follows. In Sec.~\ref{sec:kinematics} and \ref{sec:Sirlin} we briefly review the kinematics of $K_{e3}$ and Sirlin's representation of the virtual EMRC. In Sec.~\ref{sec:cancellation} we study the integrals with IR divergences and mass singularities in both the virtual corrections and bremsstrahlung process using the mass-expansion method, and prove the exact cancellation of both in the total decay rate. Next, in Sec.~\ref{sec:extra} we discuss the aforementioned contribution from the bremsstrahlung which remains finite in the $m_e\rightarrow 0$ limit but is not explicitly present in massless QED; this subtle chirality-breaking effect is crucial in reconciling the mass-expansion method and DR. In Sec.~\ref{sec:DR} we repeat a similar analysis using DR and demonstrate the equivalence between the two methods. Final discussions are given in Sec.\ref{sec:final}. 

\section{\label{sec:kinematics}Brief review of the $\boldsymbol{K_{e3}}$ kinematics}

Despite being available in many papers (e.g. Appendix A in Ref.~\cite{Seng:2021wcf}), we
still start with a brief review of the essential kinematics of the $K_{e3}$ decay to keep the discussion self-contained. We are interested in the following inclusive decay process: $K(p)\rightarrow \pi(p')+e^+(p_e)+\nu_e(p_\nu)+n\gamma$ ($n\geq 0$) with realistic physical masses, i.e. $m_e\neq 0$. If all the massless particles in the final state are left unobserved, then the decay process is fully described by three scalar kinematic variables $\{x,y,z\}$ defined as:
\begin{equation}
P^2\equiv (p-p'-p_e)^2=M_K^2x~,~p\cdot p_e=\frac{1}{2}M_K^2 y~,~p\cdot p'=\frac{1}{2}M_K^2 z~,
\end{equation}
notice that when $n=0$ we must have $x=0$ given that $P^2=p_\nu^2=0$ assuming massless neutrinos.
We also define $r_\pi\equiv M_\pi^2/M_K^2$ and $r_e\equiv m_e^2/M_K^2$ for notational simplicity. The $K$--$\pi$ squared momentum transfer is given by $t=(p-p')^2=M_K^2(1-z+r_\pi)$. Finally, in the $K^0$ ($K^+$) decay channel it is also customary to define the respective Mandelstam variables $s=(p'+p_e)^2$ and $u=(p-p_e)^2$.   

Up to $\mathcal{O}(G_F^2\alpha)$ in the decay rate, only the $n=0$ (three-body) and $n=1$ (four-body) decay processes need to be included. The corresponding decay rate formula are:
\begin{equation}
\Gamma_{3-\mathrm{body}}=\frac{M_K}{256\pi^3}\int_{\mathcal{D}_3}dydz|M|_{K\rightarrow\pi e\nu}^2~,
\end{equation}
and
\begin{eqnarray}
\Gamma_{4-\mathrm{body}}&=&\frac{M_K^3}{512\pi^4}\left\{\int_{\mathcal{D}_3}dydz\int_0^{\alpha_+}dx+\int_{\mathcal{D}_{4-3}}dydz\int_{\alpha_-}^{\alpha_+}dx\right\}\int d\Gamma_k d\Gamma_{p_\nu}\nonumber\\
&&\times (2\pi)^4\delta^{(4)}(P-k-p_\nu)|M|_{K\rightarrow\pi e\nu\gamma}^2~
\end{eqnarray}
respectively, where:
\begin{equation}
\alpha_\pm(y,z)\equiv 1-y-z+r_\pi+r_e+\frac{yz}{2}\pm\frac{1}{2}\sqrt{y^2-4r_e}\sqrt{z^2-4r_\pi}~,
\end{equation}
and we have also defined the following shorthand for the integral measure:
\begin{equation}
d\Gamma_k\equiv \frac{d^3k}{(2\pi)^3 2E_k}~.
\end{equation}
The regions $\mathcal{D}_3$ and $\mathcal{D}_{4-3}$ are defined as:
\begin{eqnarray}
\mathcal{D}_3&:&c(z)-d(z)<y<c(z)+d(z)~,\:\:2\sqrt{r_\pi}<z<1+r_\pi-r_e\nonumber\\
\mathcal{D}_{4-3}&:&2\sqrt{r_e}<y<c(z)-d(z)~,\:\:2\sqrt{r_\pi}<z<1-\sqrt{r_e}+\frac{r_\pi}{1-\sqrt{r_e}}~,
\end{eqnarray}
where
\begin{equation}
c(z)=\frac{(2-z)(1+r_e+r_\pi-z)}{2(1+r_\pi-z)}~,~d(z)=\frac{\sqrt{z^2-4r_\pi}(1+r_\pi-r_e-z)}{2(1+r_\pi-z)}~.
\end{equation}

The tree-level $n=0$ decay amplitude can be written as:
\begin{equation}
M_0=-\frac{G_F}{\sqrt{2}}L_\lambda F^\lambda(p',p)~,
\end{equation} 
where $L_\lambda=\bar{u}_\nu\gamma_\lambda(1-\gamma_5)v_e$ is the lepton current, and 
\begin{equation}
F^\lambda(p',p)=\langle \pi(p')|J_W^{\lambda\dagger}(0)|K(p)\rangle=V_{us}^*\left[f_+^{K\pi}(t)(p+p')^\lambda+f_-^{K\pi}(t)(p-p')^\lambda\right]~
\end{equation}
with $J_W^\lambda$ the charged weak current, defines the two form factors $f_\pm^{K\pi}(t)$. For simplicity we will omit the superscript $K\pi$, knowing that they refer to $K^0\pi^-$ ($K^+\pi^0$) in the $K^0$ ($K^+$)-channel.  The squared amplitude (summing over lepton spins) is given by:
\begin{equation}
|M_0|^2(x,y,z)\equiv G_F^2F_\mu(p',p)F_\nu^*(p',p)\mathrm{Tr}[\slashed{P}\gamma^\mu(\slashed{p}_e-m_e)\gamma^\nu(1-\gamma_5)]~.
\end{equation}
Here, we purposely retain the $x$-dependence so that the same structure can be reused in the bremsstrahlung process. By evaluating the trace, we find that terms involving $f_-(t)$ are suppressed by $r_e$, which makes their contribution to the decay rate negligible. Given that the virtual corrections can always be expressed as corrections to the form factors:
\begin{equation}
f_\pm(t)\rightarrow f_\pm (t)+\delta f_\pm(y,z)~,
\end{equation} 
it is therefore only $\delta f_+(y,z)$ which is of relevance in practise.

\section{\label{sec:Sirlin}Sirlin's representation of the long-distance EMRC}

\begin{figure}[tb]
	\begin{centering}
		\includegraphics[scale=0.15]{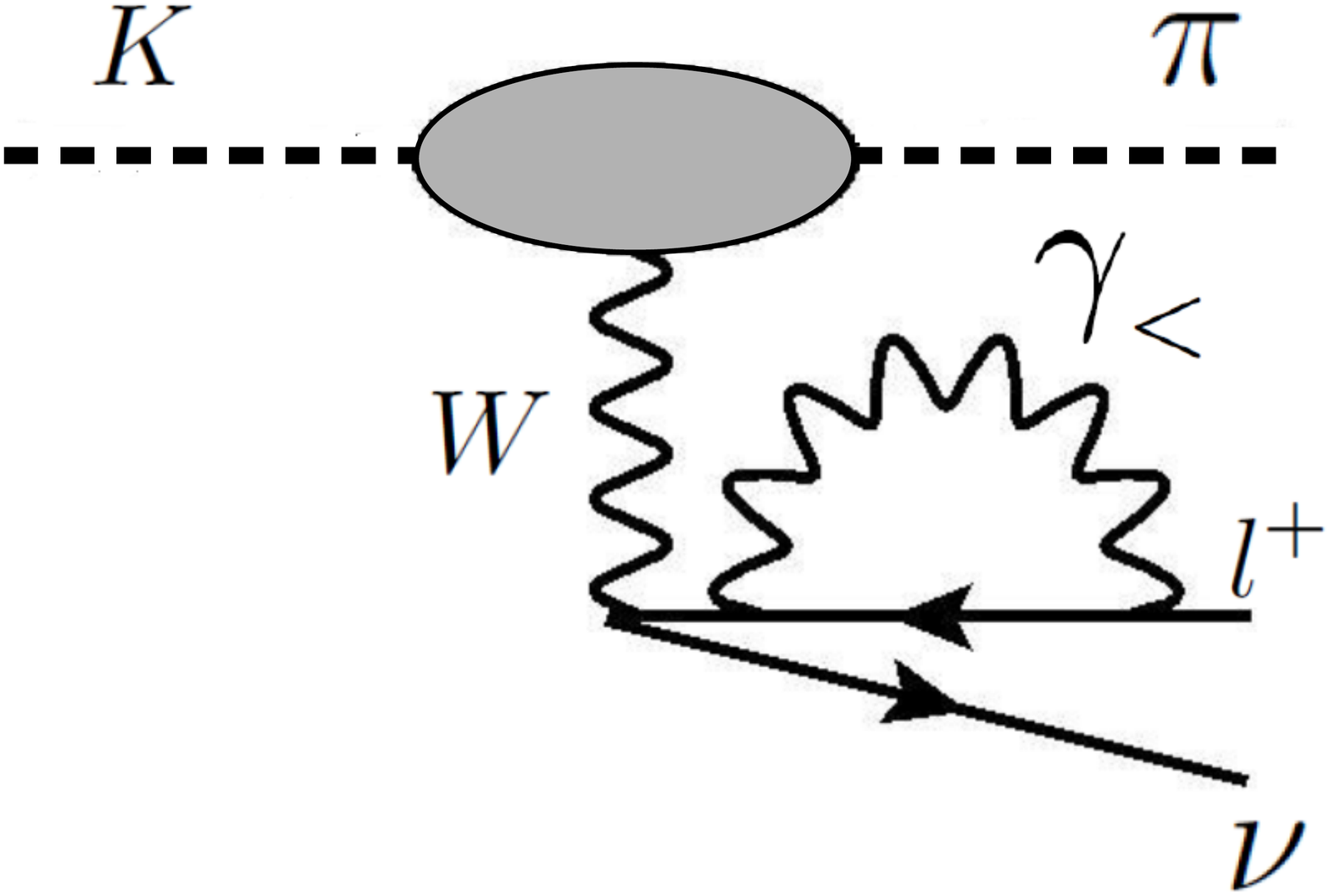}
		\includegraphics[scale=0.15]{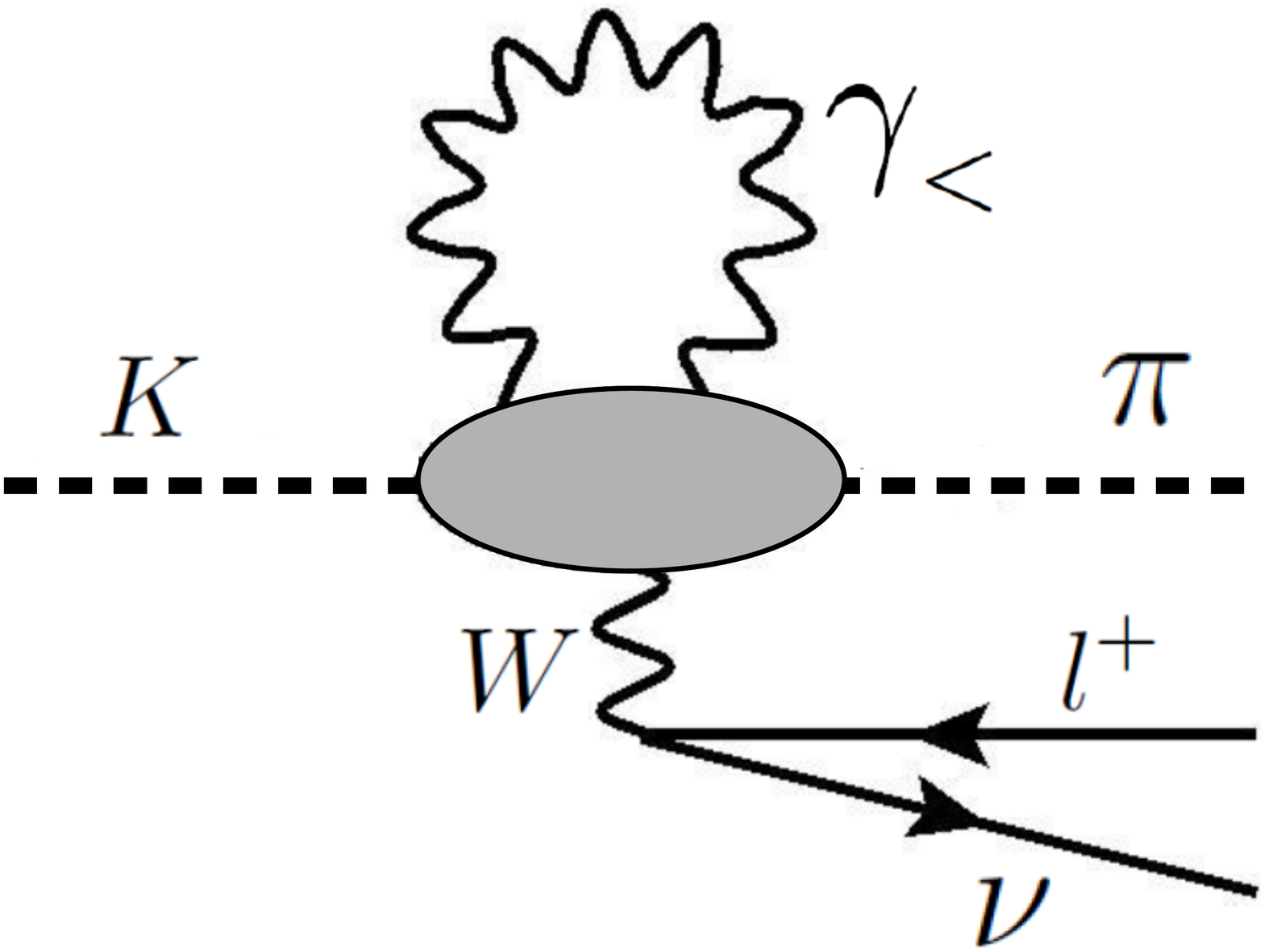}
		\includegraphics[scale=0.15]{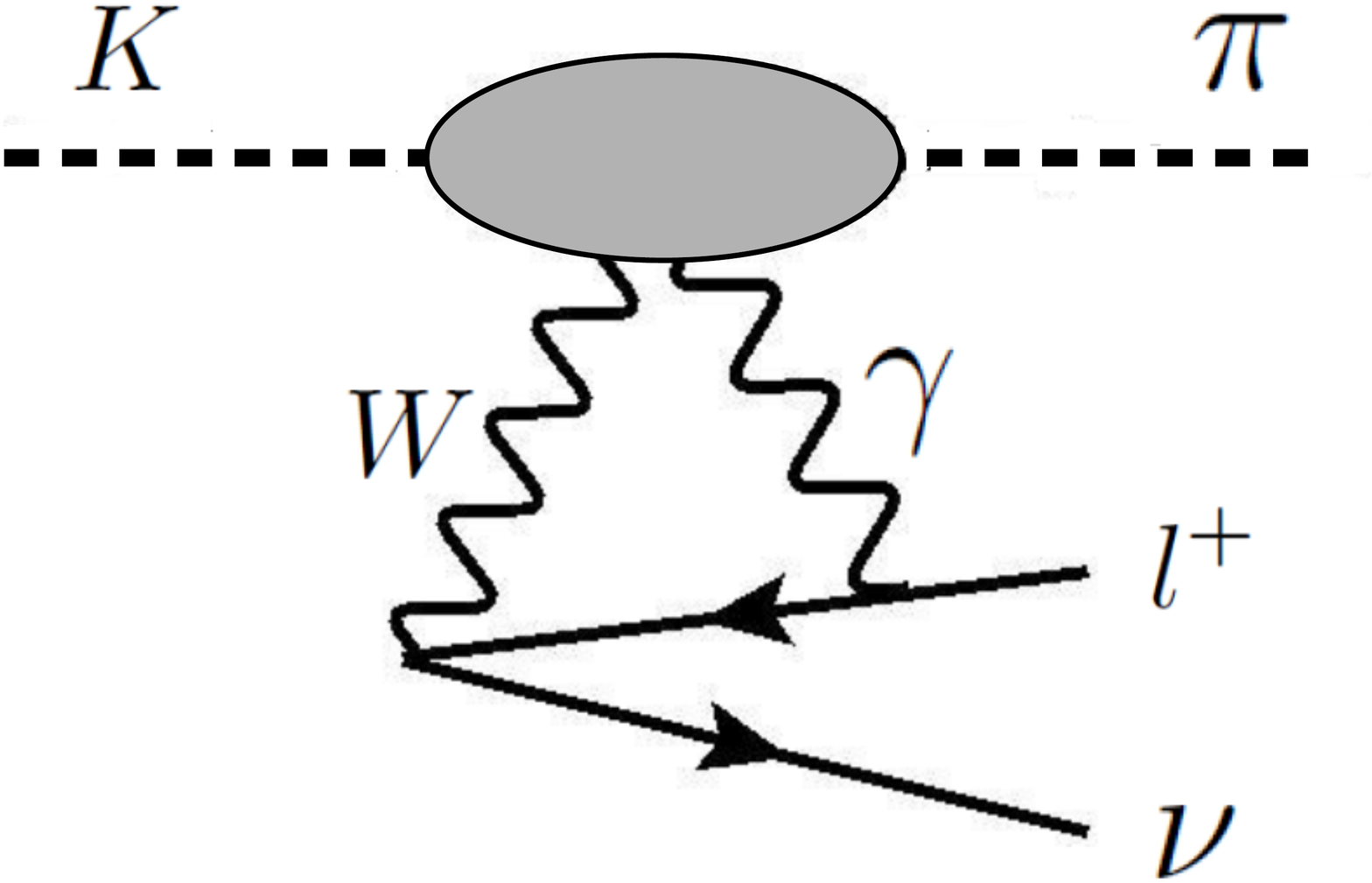}
		\par\end{centering}
		\caption{\label{fig:EMdiagrams}One-loop EMRC in Sirlin's representation. $\gamma_<$ denotes a photon which propagator is attached to a Pauli-Viilars-regulator $M_W^2/(M_W^2-q^{\prime 2})$.}
\end{figure}

In Sirlin's representation (Ref.~\cite{Sirlin:1977sv}, see also Refs.~\cite{Seng:2019lxf,Seng:2020jtz,Seng:2021syx} for comprehensive reviews), the full $\mathcal{O}(\alpha)$ virtual electroweak RC splits into two parts. First is the ``weak'' RC that depend only on physics at the scale $M_W$; they come from one-loop diagrams that involve only heavy gauge bosons, or with photon but only pick up the first term in the following splitting of the photon propagator:
\begin{equation}
\frac{-ig_{\mu\nu}}{q^{\prime 2}-M_\gamma^2}=\frac{-ig_{\mu\nu}}{q^{\prime 2}-M_W^2}+\frac{M_W^2}{M_W^2-q^{\prime 2}}\frac{-ig_{\mu\nu}}{q^{\prime 2}-M_\gamma^2}+\mathcal{O}(M_\gamma^2)~.    
\end{equation}
The weak RC are either reabsorbed into the definition of the Fermi constant $G_F$ or provides a regular correction that is proportional to the tree-level amplitude which is irrelevant for our discussion. What remains are the long-distance EMRC represented by the three Feynman diagrams in Fig.\ref{fig:EMdiagrams}:
\begin{enumerate}
	\item Contribution from the electron wavefunction renormalization: $\delta M=(1/2)\delta Z_e M_0$,
	\item The long-distance EMRC to the $K\pi$ form factor, which can be divided into ``two-point function'' and ``three-point function'': $\delta M=\delta M_2+\delta M_3$,
	\item The $\gamma W$-box diagram, which splits into the piece $\delta M_{\gamma W}=\delta M_{\gamma W}^a+\delta M_{\gamma W}^b$ without (a) and with (b) a totally-antisymmetric tensor.
\end{enumerate}
Furthermore, using Ward identities, one may combine $\delta M_2$ and $\delta M_{\gamma W}^a$ to get an ``analytic'' piece and a ``residual integral'' piece:
\begin{equation}
\delta M_2+\delta M_{\gamma W}^a=(\delta M_2+\delta M_{\gamma W}^a)_\mathrm{ana}+(\delta M_2+\delta M_{\gamma W}^a)_\mathrm{int}~.
\end{equation}
Throughout this work we ignore small corrections from perturbative QCD that are relevant to the actual numerical calculation at the level of $10^{-4}$~\cite{Sirlin:1977sv} but do not affect the structure of the infrared and mass singularities. This gives:
\begin{eqnarray}
(\delta M_2+\delta M_{\gamma W}^a)_\mathrm{ana}&=&-\left(\frac{\alpha}{8\pi}+2ie^2\int\frac{d^4q'}{(2\pi)^4}\frac{M_W^2}{M_W^2-q^{\prime 2}}\frac{1}{[(p_e-q')^2-m_e^2]q^{\prime 2}}\right)M_0\nonumber\\
(\delta M_2+\delta M_{\gamma W}^a)_\mathrm{int}&=&\frac{G_Fe^2}{\sqrt{2}}L_\lambda\int\frac{d^4q'}{(2\pi)^4}\frac{1}{(p_e-q')^2-m_e^2}\biggl\{\frac{2p_e\cdot q'q^{\prime\lambda}}{(q^{\prime 2}-M_\gamma^2)^2}T^\mu_{\:\:\mu}\nonumber\\
&&+\frac{2p_{e\mu}}{q^{\prime 2}-M_\gamma^2}T^{\mu\lambda}-\frac{(p-p')_\mu}{q^{\prime 2}-M_\gamma^2}T^{\lambda\mu}+\frac{i}{q^{\prime 2}-M_\gamma^2}\Gamma^\lambda\biggr\}~.\label{eq:Sirlin1}
\end{eqnarray}
The antisymmetric piece of the $\gamma W$-box diagram reads:
\begin{equation}
\delta M_{\gamma W}^{b}=-i\frac{G_Fe^2}{\sqrt{2}}L_\lambda\int\frac{d^4q'}{(2\pi)^4}\frac{M_W^2}{M_W^2-q^{\prime 2}}\frac{1}{(p_e-q')^2-m_e^2}\frac{1}{q^{\prime 2}}\epsilon^{\mu\nu\alpha\lambda}q_\alpha'T_{\mu\nu}~.\label{eq:Sirlin2}
\end{equation}
The non-trivial integrals above are expressed in terms of the following quantities:
\begin{eqnarray}
T^{\mu\nu}(q';p',p)&\equiv &\int d^4x e^{iq'\cdot x}\langle \pi(p')|T\{J_\mathrm{em}^\mu(x)J_W^{\nu\dagger}(0)\}|K(p)\rangle~,\nonumber\\
\Gamma^\mu(q';p',p)&\equiv&\int d^4xe^{iq'\cdot x}\langle\pi(p')|T\{J_\mathrm{em}^\mu(x)\partial\cdot J_W^\dagger(0)\}|K(p)\rangle~.
\end{eqnarray}

\section{\label{sec:cancellation}Cancellation of IR and mass singularities}

To check the cancellation of IR  and mass singularities we only need the divergent pieces from the virtual and real corrections; but to prove the equivalence between the mass-expansion method and DR we need the finite pieces as well. Fortunately, not all the finite pieces are relevant: there are parts of the loop/bremsstrahlung integrals that are intrinsically finite even at $M_\gamma=m_e=0$, so these parts must be the same in both methods and can be discarded in our discussion. What we are interested in are the loop/bremsstrahlung integrals with IR and mass singularities, which we always split 
into divergent and finite pieces:
\begin{equation}
F=F^\mathrm{div,m}+F^\mathrm{fin}+\mathcal{O}(M_\gamma^2, m_e^2)~,\label{eq:Fmass}
\end{equation}
where $F^\mathrm{div,m}$ (``m'' stands for ``mass-expansion'') contains only the terms proportional to $\ln (m_e^2/\mu^2)$ or $\ln(M_\gamma^2/\mu^2)$ (or both), with $\mu$ an arbitrarily chosen scale, while $F^\mathrm{fin}$ represents all the remaining finite terms independent of $M_\gamma$ and $m_e$. The sum of the two pieces is of course $\mu$-independent. Analytic formulas for these integrals are in given in Ref.~\cite{Seng:2021wcf} at non-zero $m_e$; here, we just need to expand those expression with respect to $m_e$. Notice that this procedure implicitly assumes the hierarchy $m_e\gg M_\gamma$ in the mass-expansion method.  

Throughout this study, we work in the Feynman gauge.
We start from the electron wavefunction renormalization $\delta Z_e=\delta Z_e^\mathrm{div,m}+\delta Z_e^\mathrm{fin}$,
where
\begin{equation}
\delta Z_e^\mathrm{div,m}=-\frac{\alpha}{4\pi}\left[-3\ln\frac{m_e^2}{\mu^2}+2\ln\frac{M_\gamma^2}{\mu^2}\right]~,~\delta Z_e^\mathrm{fin}=-\frac{\alpha}{4\pi}\left[\ln\frac{M_W^2}{\mu^2}+\frac{9}{2}\right]~.
\end{equation}
Next, the ``analytic'' piece in $\delta M_2+\delta M_{\gamma W}^a$ gives:
\begin{equation}
(\delta M_2+\delta M_{\gamma W}^a)_\mathrm{ana}^\mathrm{div,m}=-\frac{\alpha}{2\pi}\ln\frac{m_e^2}{\mu^2}M_0~,~(\delta M_2+\delta M_{\gamma W}^a)_\mathrm{ana}^\mathrm{fin}=\frac{\alpha}{2\pi}\left(\ln\frac{M_W^2}{\mu^2}+\frac{3}{4}\right)M_0~.
\end{equation}
These are all exactly-known contributions independent of hadronic structure. 

\subsection{\label{sec:Born}Born contribution to the remaining loop integrals}

Evaluating the remaining loop integrals $(\delta M_2+\delta M_{\gamma W}^a)_\mathrm{int}$ and $\delta M_{\gamma W}^b$ in the virtual correction requires the knowledge of the hadronic tensor $T^{\mu\nu}$ and the
vertex function $\Gamma^\mu$. A particularly important set of constraints comes from the EM and charged weak Ward identities:
\begin{eqnarray}
\text{EM}&:& q_\mu' T^{\mu\nu}(q';p',p)=-iF^\nu(p',p)~,\nonumber\\
\text{Charged weak}&:&q_\nu T^{\mu\nu}(q';p',p)=-iF^\mu(p',p)-i\Gamma^\mu(q';p',p)~,
\end{eqnarray}
where $q=p'-p+q'$. We have used both in the derivation of Eqs.\eqref{eq:Sirlin1}, \eqref{eq:Sirlin2}.

In Refs.~\cite{Seng:2021boy,Seng:2021wcf,Seng:2022wcw}, the so-called ``convection term contribution''~\cite{Meister:1963zz} was analytically calculated. It involves the following substitution for the hadronic tensor
and the vertex function:
\begin{equation}
T_{K^0\pi^-}^{\mu\nu,\mathrm{conv}}=-\frac{i(2p'+q')^\mu F^\nu(p',p)}{(p'+q')^2-M_\pi^2}~,~
T_{K^+\pi^0}^{\mu\nu,\mathrm{conv}}=\frac{i(2p-q')^\mu F^\nu(p',p)}{(p-q')^2-M_K^2}~,
\end{equation}
and:
\begin{equation}
\Gamma^{\mu,\mathrm{conv}}_{K^0\pi^-}=\frac{(2p'+q')^\mu (p'-p)_\lambda F^\lambda(p',p)}{(p'+q')^2-M_\pi^2}~,
~\Gamma^{\mu,\mathrm{conv}}_{K^+\pi^0}=-\frac{(2p-q')^\mu(p'-p)_\lambda F^\lambda(p',p)}{(p-q')^2-M_K^2}~.
\end{equation}
Those relationships represent the simplest structures that satisfy the exact EM Ward identity and are able to reproduce the full IR divergence structure. However, they fail to satisfy the charged weak Ward identity even in the SU(3)$_f$ limit, and therefore do not reproduce the full $\ln m_e$ structure. 

Fortunately, in the actual non-perturbative numerical calculation of those works, what enters is not just the convection term, but the full ``Born'' contribution which includes more complete pole and seagull structures. Its original implementation in the numerical calculation can be found in Eqs.(4.3)-(4.5) of Ref.~\cite{Seng:2021wcf}, but there the $K\pi$ form factors depend on the loop momentum $q'$, which complicates its analytic study. In this paper we make a further simplification, namely to discard the $q'$-dependence in the electromagnetic and the $f_\pm$ form factors, which is again relevant
to the actual numerical calculation but do not affect the divergent structure:
\begin{eqnarray}
T_{K^0\pi^-}^{\mu\nu,\mathrm{B}}&=&-iV_{us}^*\left[\frac{(2p'+q')^\mu}{(p'+q')^2-M_\pi^2}\left(f_+(t)(2p+q)^\nu-f_-(t)q^\nu\right)-g^{\mu\nu}\left(f_+(t)-f_-(t)\right)\right],\nonumber\\
T_{K^+\pi^0}^{\mu\nu,\mathrm{B}}&=&iV_{us}^*\left[\frac{(2p-q')^\mu}{(p-q')^2-M_K^2}\left(f_+(t)(2p'-q)^\nu-f_-(t)q^\nu\right)-g^{\mu\nu}\left(f_+(t)+f_-(t)\right)\right],\nonumber\\
\Gamma^{\mu,\mathrm{B}}_{K^0\pi^-}&=&-V_{us}^*\frac{M_K^2-M_\pi^2}{(p'+q')^2-M_\pi^2}(2p'+q')^\mu\left(f_+(t)+\frac{q^2}{M_K^2-M_\pi^2}f_-(t)\right),\nonumber\\
\Gamma^{\mu,\mathrm{B}}_{K^+\pi^0}&=&V_{us}^*\frac{M_K^2-M_\pi^2}{(p-q')^2-M_K^2}(2p-q')^\mu\left(f_+(t)+\frac{q^2}{M_K^2-M_\pi^2}f_-(t)\right)~.
\end{eqnarray}
They satisfy the following relations:
\begin{eqnarray}
q_\mu'T^{\mu\nu,\mathrm{B}}(q';p',p)+iF^\nu(p',p)&=&0~,\nonumber\\
q_\nu T^{\mu\nu,\mathrm{B}}(q';p',p)+iF^\mu(p',p)+i\Gamma^{\mu,\mathrm{B}}(q';p',p)&=&iV_{us}^*f_-(t)\left(2(p-p')^\mu-q^{\prime\mu}\right)~,
\end{eqnarray}
so the EM Ward identity is obeyed. Meanwhile, the charged weak Ward identity is not exactly satisfied. There is an extra pole-free term on the right hand side of the second line. Fortunately, this term has only $(p-p')^\mu$ and $q^{\prime \mu}$ structures, which means: upon plugging into the loop integrals, it can only give rise to $\delta f_-$ and not $\delta f_+$. So, as far as its contribution to $\delta f_+$ is concerned, the Born term satisfies both Ward identities. Therefore, it is able to describe the complete mass singularities in the virtual correction within the framework of Sirlin's representation. 

Now, we can discuss the divergence structures in the remaining integrals. First, as explained in Ref.~\cite{Seng:2021wcf}, the  IR divergences from $\delta M_2$ (contained in the first term in $(\delta M_2+\delta M_{\gamma W}^a)_\mathrm{int}$) combined with $\delta M_3$ gives:
\begin{equation}
(\delta f_+)_\mathrm{2pt+3pt}^\mathrm{div,m}=-\frac{\alpha}{4\pi}\ln\frac{M_\gamma^2}{\mu^2}f_+~.\label{eq:2pt+3pt}
\end{equation}
Its finite contribution is irrelevant for our discussion and need not be retained. Next, the second term in $(\delta M_2+\delta M_{\gamma W}^a)_\mathrm{int}$ contains both IR and mass singularities, which comes from the following scalar function~\footnote{Notice that the definition of $C_0^\mathrm{fin}$ here and $I_i^\mathrm{fin}$ in Eq.\eqref{eq:Ii} are different from those in Ref.~\cite{Seng:2021wcf}, because the full results there were not expanded with respect to $m_e$.}:
\begin{equation}
C_0\equiv\frac{16\pi^2}{i}\int\frac{d^4k}{(2\pi)^4}\frac{1}{[(p_1-k)^2-M_1^2][(p_e-k)^2-m_e^2][k^2-M_\gamma^2]}=C_0^\mathrm{div,m}+C_0^\mathrm{fin}
\end{equation}
where
\begin{eqnarray}
C_0^\mathrm{div,m}&=&\frac{1}{M_1^2-v}\left\{-\frac{1}{2}\ln\frac{m_e^2}{\mu^2}\ln\frac{M_\gamma^2}{\mu^2}+\frac{1}{4}\ln^2\frac{m_e^2}{\mu^2}+\ln\frac{M_\gamma^2}{\mu^2}\left[\frac{1}{2}\ln\frac{M_1^2}{\mu^2}-\ln\frac{M_1^2}{M_1^2-v}\right]\right\}\nonumber\\
C_0^\mathrm{fin}&=&\frac{1}{M_1^2-v}\left\{-\frac{1}{4}\ln^2\frac{M_1^2}{\mu^2}+\ln\frac{M_1^2}{\mu^2}\ln\frac{M_1^2}{M_1^2-v}-\frac{1}{2}\ln^2\frac{M_1^2}{M_1^2-v}+\mathrm{Li}_2\left(\frac{v}{v-M_1^2}\right)\right\}~.\nonumber\\
\end{eqnarray}
Here we should take $M_1=M_\pi$, $v=s$ in the $K^0$-decay, and $M_1=M_K$, $v=u$ in the $K^+$-decay.
Finally, the mass singularities from the third and fourth terms in $(\delta M_2+\delta M_{\gamma W}^a)_\mathrm{int}$ cancel out with that from $\delta M_{\gamma W}^b$, resulting in a finite sum. (If only the convection term contribution is considered,  the cancellation does not occur and one obtains an extra $\ln m_e$-divergent piece from $\delta M_{\gamma W}^b$). Finally, the sum of the divergent contributions to the virtual corrections reads:
\begin{eqnarray}
(\delta f_+)^\mathrm{div,m}&=&\left\{\frac{1}{2}\delta Z_e^\mathrm{div,m}-\frac{\alpha}{2\pi}(v-M_i^2)C_0^\mathrm{div,m}-\frac{\alpha}{4\pi}\ln\frac{M_\gamma^2}{\mu^2}\right\}f_+
\nonumber\\&=&-\frac{\alpha}{4\pi}\left\{\ln\frac{m_e^2}{\mu^2}\ln\frac{M_\gamma^2}{\mu^2}-\frac{1}{2}\ln^2\frac{m_e^2}{\mu^2}+\frac{1}{2}\ln\frac{m_e^2}{\mu^2}+\ln\frac{M_\gamma^2}{\mu^2}\left(2-\ln\frac{M_i^2}{\mu^2}+2\ln\frac{M_i^2}{M_i^2-v}\right)\right\}f_+\nonumber\\\label{eq:deltafpmass}
\end{eqnarray}
Throughout this work we use ``$i$'' to label the charged meson in the decay process, i.e. $\pi^-$ ($K^+$) in the $K^0$ ($K^+$)-decay. 

\subsection{Divergent integrals in bremsstrahlung}

\begin{figure}[tb]
	\begin{centering}
		\includegraphics[scale=0.4]{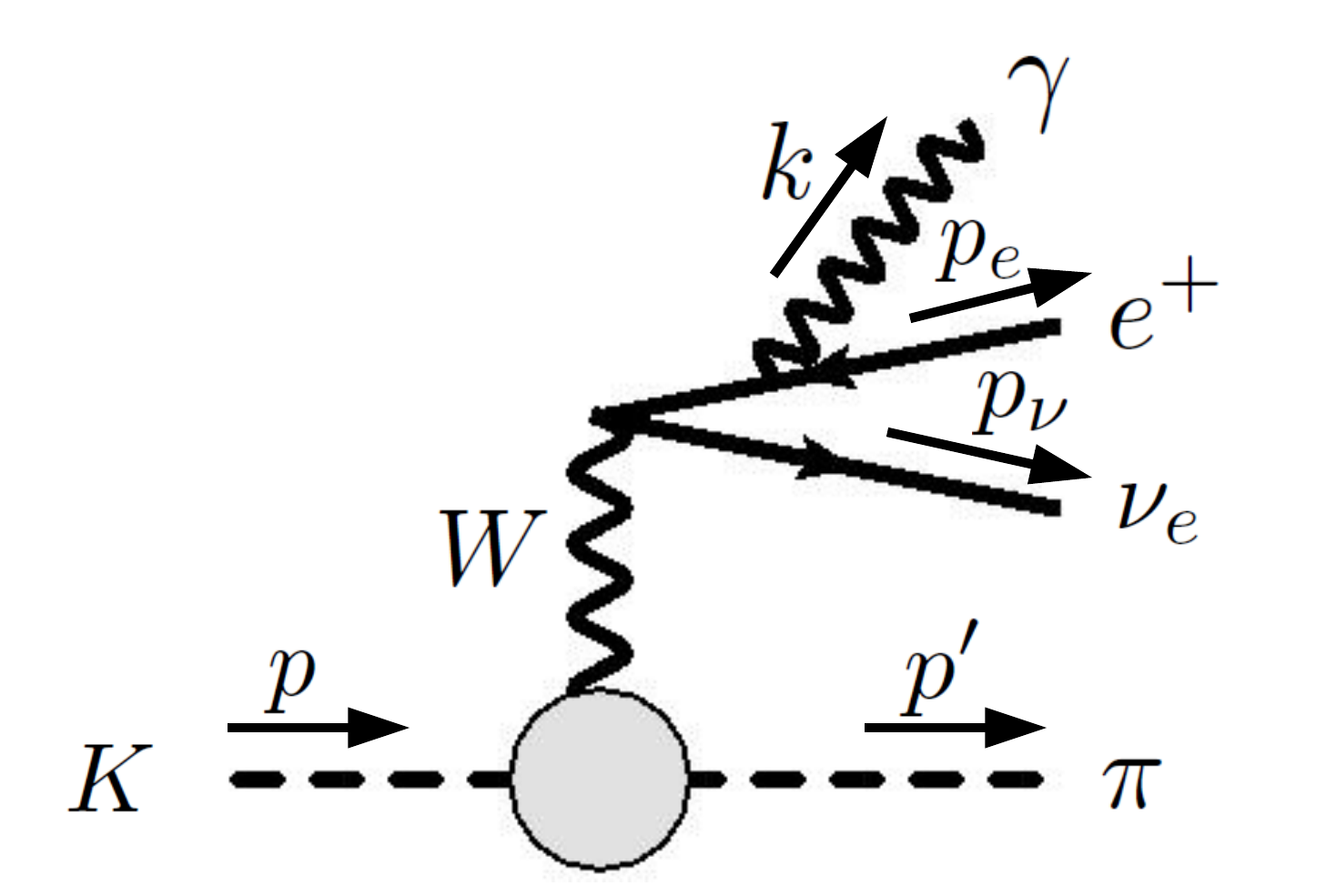}
		\includegraphics[scale=0.4]{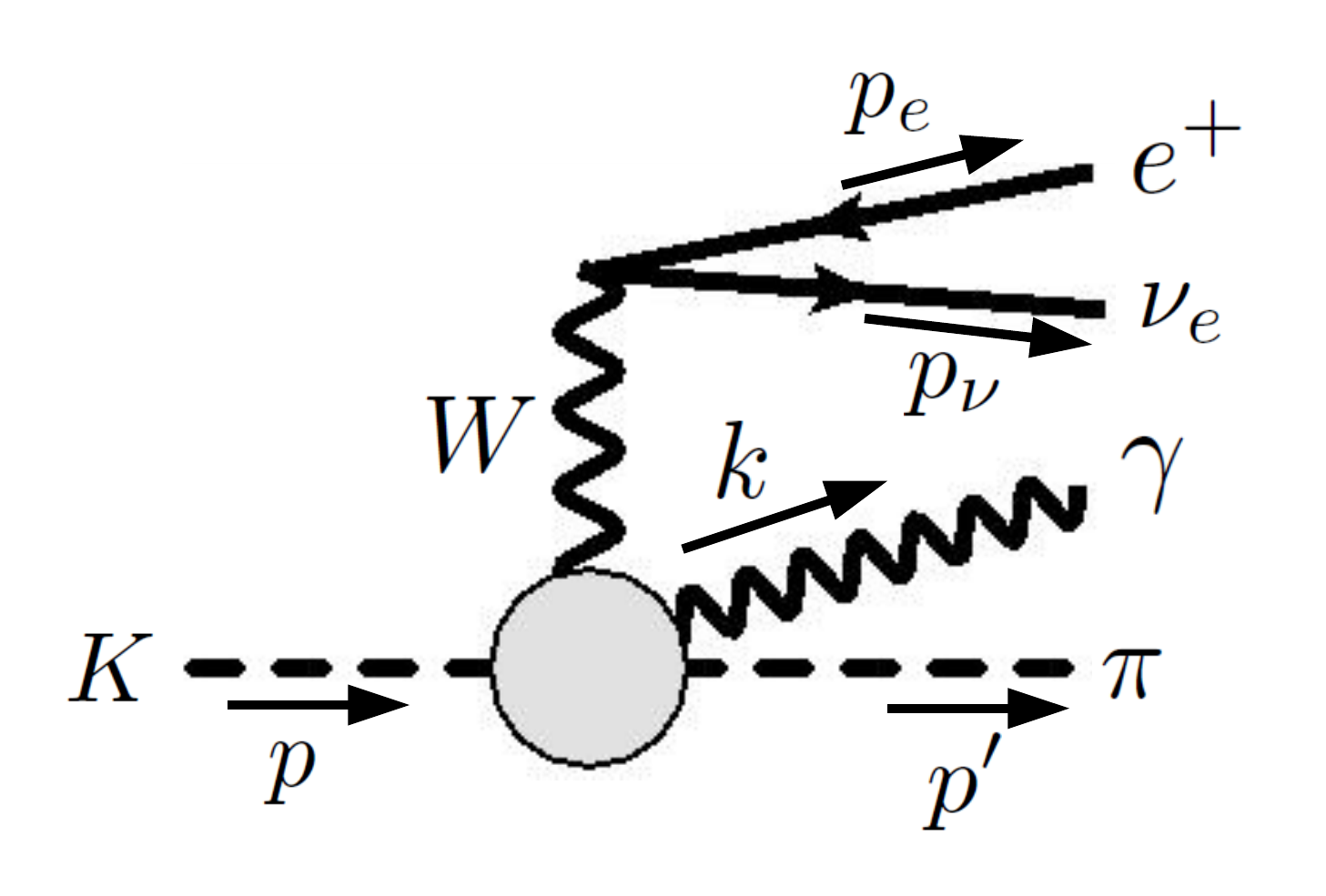}
		\par\end{centering}
	\caption{\label{fig:brems}Tree-level bremsstrahlung diagrams contributing to the $K_{e3}$ decay rate.}
\end{figure}

Next we study the IR and mass singularities from the bremsstrahlung process depicted in Fig.\ref{fig:brems}. Following the treatment in Refs.~\cite{Seng:2021boy,Seng:2021wcf,Seng:2022wcw}, we split the bremsstrahlung amplitude as:
\begin{equation}
M_{K\rightarrow\pi e\nu\gamma}=M_A+M_B~,
\end{equation}
where $M_A$ contains the full convection term contribution:
\begin{equation}
M_A=-\frac{eG_F}{\sqrt{2}}F_\mu(p',p)\varepsilon_\nu^*(k)\bar{u}_\nu\gamma^\mu(1-\gamma_5)\left\{\left(\frac{p_e}{p_e\cdot k}-\frac{p_i}{p_i\cdot k}\right)^\nu+\frac{1}{2p_e\cdot k}\slashed{k}\gamma^\nu\right\}v_e~,
\end{equation}
and $M_B$ is the remaining regular term that admits a ChPT expansion.
It is easy to check that all the singularities exist only in the phase space integral over $|M_A|^2$. In particular, the IR divergence exists only in the $\mathcal{D}_3$ region, which can be isolated through the following separation:
\begin{equation}
|M_A|^2=-e^2\left(\frac{p_e}{p_e\cdot k}-\frac{p_i}{p_i\cdot k}\right)^2|M_0|^2(0,y,z)+|M_A|_\mathrm{res}^2~.
\end{equation}
The integral of the first term gives the full IR divergence (and a part of the $\ln m_e$ divergences), while the second term gives only $\ln m_e$ but not IR divergences. Meanwhile, the integral of the whole $|M_A|^2$ in the $\mathcal{D}_{4-3}$ region gives the last piece with mass singularities. Therefore, for the purpose of this paper only these three contributions need to be studied. 

First, the IR divergence is fully contained in the following integral:
\begin{equation}
I_i\equiv \int_0^{\alpha_+}dx\int d\Gamma_k d\Gamma_{p_\nu}(2\pi)^4\delta^{(4)}(P-k-p_\nu)\left(\frac{p_i}{p_i\cdot k}-\frac{p_e}{p_e\cdot k}\right)^2=I_i^\mathrm{div,m}+I_i^\mathrm{fin}~.\label{eq:Ii}
\end{equation}
Its analytic expression is given in Appendix D of Ref.~\cite{Seng:2021wcf}, and here we further expand it with respect to $m_e$. This gives:
\begin{eqnarray}
I_i^\mathrm{div,m}&=&\frac{1}{4\pi M_K^2}\biggl\{-\ln\frac{m_e^2}{\mu^2}\ln\frac{M_\gamma^2}{\mu^2}+\frac{1}{2}\ln^2\frac{m_e^2}{\mu^2}+\left[1+2\ln\left(\frac{\alpha_+}{1-z+r_\pi}\right)\right]\ln\frac{m_e^2}{\mu^2}\nonumber\\
&&-\left[2+\ln\left(\frac{\mu^2P_0^2(0)(1-\cos\chi)^2}{M_K^4(1-z+r_\pi)^2}\right)\right]\ln\frac{M_\gamma^2}{\mu^2}\biggr\}\nonumber\\
I_i^\mathrm{fin}&=&\frac{1}{4\pi M_K^2}\biggl\{-\frac{1}{2}\ln^2\left(\frac{\mu^2P_0^2(0)(1-\cos\chi)^2}{M_K^4\alpha_+^2}\right)+4\ln^2\left(\sin\frac{\chi}{2}\right)+\ln\left(\frac{M_K^4\alpha_+^2}{4P_0^2(0)\mu^2}\right)\nonumber\\
&&+2\mathrm{Li}_2\left(\cos^2\frac{\chi}{2}\right)+2\ln^2\left(\frac{\alpha_+}{1-z+r_\pi}\right)+2\ln\left(\frac{\alpha_+}{1-z+r_\pi}\right)+4\mathrm{Li}_2\left(\frac{\alpha_+}{1-z+r_\pi}\right)\biggr\}\nonumber\\
&&+\frac{1}{\pi M_K^2}\int_0^{\alpha_+}dx\frac{1}{x}\ln\left(\frac{P_0(x)-P_1(x)}{P_0(0)-P_1(0)}\right)~.
\end{eqnarray}
where we have taken the $m_e=0$ kinematics:
\begin{equation}
P_0(x)=\frac{p_i\cdot P}{M_i}~, ~P_0(x)-P_1(x)=\frac{p_e\cdot P}{p_i\cdot p_e}M_i~,~\chi\equiv\cos^{-1}\frac{P_1(0)}{P_0(0)}~.
\end{equation}

The remaining bremsstrahlung integral with mass singularities is:
\begin{equation}
I_{m,1}(p_i)\equiv 8\pi\int d\Gamma_k d\Gamma_{p_\nu}\frac{(2\pi)^4\delta^{(4)}(P-k-p_\nu)}{(p_i\cdot k)^mp_e\cdot k}=I_{m,1}^\mathrm{div,m}(p_i)+I_{m,1}^\mathrm{fin}(p_i)~,\label{eq:Im1mass}
\end{equation}
with $m=1,0,-1,-2$. The divergent piece of these integrals is:
\begin{equation}
I_{m,1}^\mathrm{div,m}(p_i)=-\frac{2(2p_e\cdot P)^{m-1}}{(M_K^2 x)^m(p_i\cdot p_e)^m}\ln\frac{m_e^2}{\mu^2}~,
\end{equation} 
while the finite pieces are:
\begin{eqnarray}
I_{1,1}^\mathrm{fin}(p_i)&=&-\frac{2}{ p_e\cdot p_i M_K^2x}\ln\left(\frac{\mu^2 M_i^2}{4(p_e\cdot p_i)^2}\right)~,\nonumber\\
I_{0,1}^\mathrm{fin}(p_i)&=&-\frac{1}{ p_e\cdot P}\ln\left(\frac{\mu^2 M_K^2 x}{4(p_e\cdot P)^2}\right)~,\nonumber\\
I_{-1,1}^\mathrm{fin}(p_i)&=&-\frac{p_i\cdot p_e}{2 (p_e\cdot P)^2}M_K^2x\left[2+\ln\left(\frac{\mu^2 M_K^2 x}{4(p_e\cdot P)^2}\right)\right]+\frac{p_i\cdot P}{p_e\cdot P}~,\nonumber\\
I_{-2,1}^\mathrm{fin}(p_i)&=&-\frac{(p_e\cdot p_i)^2}{4 (p_e\cdot P)^3}M_K^4x^2\left[3+\ln\left(\frac{\mu^2 M_K^2 x}{4(p_e\cdot P)^2}\right)\right]\nonumber\\
&&+\frac{1}{4 p_e\cdot P}\left[\frac{2p_i\cdot P p_i\cdot p_e}{p_e\cdot P}M_K^2x-M_i^2M_K^2x+2(p_i\cdot P)^2\right]~.
\end{eqnarray}

These are all the divergent integrals that appear in the bremsstrahlung.

\subsection{\label{sec:cancel}Full cancellation of the IR and mass singularities}

Now we are ready to demonstrate the complete cancellation of IR divergences and electron mass singularities in the radiative inclusive total $K_{e3}$ decay rate. Since in the previous subsections we explicitly isolated all the terms that are logarithmically divergent with respect to $M_\gamma$ and $m_e$, in what follows we will take $M_\gamma=m_e=0$ everywhere except in those logs. We start with the contribution from $(\delta f_+)^\mathrm{div,m}$:
\begin{eqnarray}
(\delta \Gamma)_{(\delta f_+)^\mathrm{div,m}}&=&\frac{M_K}{256\pi^3}\int_{\mathcal{D}_3}dydz\left\{\frac{2\mathfrak{Re}(\delta f_+)^\mathrm{div,m}}{f_+}|M_0|^2(0,y,z)\right\}\nonumber\\
&=&\frac{\alpha G_F^2|V_{us}|^2M_K^5}{64\pi^4}\int_{2\sqrt{r_\pi}}^{1+r_\pi}dzf_+^2(t)\int_{c(z)-d(z)}^{c(z)+d(z)}dy\biggl[\ln\frac{m_e^2}{\mu^2}\ln\frac{M_\gamma^2}{\mu^2}-\frac{1}{2}\ln^2\frac{m_e^2}{\mu^2}+\frac{1}{2}\ln\frac{m_e^2}{\mu^2}\nonumber\\
&&+\ln\frac{M_\gamma^2}{\mu^2}\left(2-\ln\frac{M_i^2}{\mu^2}+\ln\frac{M_i^4}{(M_i^2-v)^2}\right)\biggr][r_\pi+(y-1)(y+z-1)]~.
\end{eqnarray}

Next, the divergent part of the bremsstrahlung integral $I_i$ contributes:
\begin{eqnarray}
(\delta \Gamma)_{I_i^\mathrm{div,m}}&=&\frac{M_K^3}{512\pi^4}\int_{\mathcal{D}_3}dydz(-e^2)I_i^\mathrm{div,m}|M_0|^2(0,y,z)\nonumber\\
&=&\frac{\alpha G_F^2|V_{us}|^2M_K^5}{64\pi^4}\int_{2\sqrt{r_\pi}}^{1+r_\pi}dzf_+^2(t)\int_{c(z)-d(z)}^{c(z)+d(z)}dy\biggl[-\ln\frac{m_e^2}{\mu^2}\ln\frac{M_\gamma^2}{\mu^2}+\frac{1}{2}\ln^2\frac{m_e^2}{\mu^2}\nonumber\\
&&+\ln\frac{m_e^2}{\mu^2}\left(1+\ln\frac{\alpha_+^2}{(1-z+r_\pi)^2}\right)-\ln\frac{M_\gamma^2}{\mu^2}\left(2-\ln\frac{M_i^2}{\mu^2}+\ln\frac{M_i^4}{(M_i^2-v)^2}\right)\biggr]~.\nonumber\\
&&\times[r_\pi+(y-1)(y+z-1)]~.
\end{eqnarray}

Summing up the above two terms, the IR divergences completely cancel, as originally demonstrated in Ref.~\cite{Seng:2021wcf}. However, the $\ln(m_e^2/\mu^2)$  terms do not fully cancel; they must be combined with those from $|M_A|_\mathrm{res}^2$ in the $\mathcal{D}_3$ region and from $|M_A|^2$ in the $\mathcal{D}_{4-3}$ region. To include these contributions, we first write:
\begin{equation}
|M_A|^2_\mathrm{res}=\sum_{m,n}\frac{\mathbb{C}_{m,n}(x,y,z)}{(p_i\cdot k)^m(p_e\cdot k)^n}~,~|M_A|^2=\sum_{m,n}\frac{\mathbb{C}'_{m,n}(x,y,z)}{(p_i\cdot k)^m(p_e\cdot k)^n}~,
\end{equation}
where the coefficients $\mathbb{C}_{m,n}^{(\prime)}(x,y,z)$ are known quantities. We are only interested in the coefficients with $n=1$ as they are attached to the $\ln m_e$-divergent integrals $I_{m,1}$. To that end, it is useful to define a new set of coefficients 
\begin{equation}
\sum_m \mathbb{C}_{m1}^{(\prime)}(x,y,z)\left\{-\frac{2(2p_e\cdot P)^{m-1}}{(M_K^2x)^m(p_i\cdot p_e)^m}\right\}\equiv 64\pi\alpha M_K^2G_F^2|V_{us}|^2 f_+^2(t)\bar{\mathbb{C}}^{(\prime)}(x,y,z)~,
\end{equation}
which turns out to be channel-independent:
\begin{eqnarray}
\bar{\mathbb{C}}(x,y,z)&=&-\frac{x^2y^2}{(1-x-z+r_\pi)^3}+\frac{xy(r_\pi-x+3y-1)}{(1-x-z+r_\pi)^2}+\frac{2y(r_\pi+2y-1)-3xy+x}{1-x-z+r_\pi}-2y,\nonumber\\
\bar{\mathbb{C}}'(x,y,z)&=&\bar{\mathbb{C}}(x,y,z)+\frac{2(r_\pi+(y-1)(y+z-1))}{x}~.
\end{eqnarray}
Based on the above, the final $\ln m_e$ singularities 
contribution from bremsstrahlung becomes:
\begin{eqnarray}
(\delta \Gamma)_{I_{m,1}^\mathrm{div,m}}&=&\frac{M_K^3}{4096\pi^5}\sum_m\left\{\int_{\mathcal{D}_3}dydz\int_0^{\alpha_+} dx\mathbb{C}_{m,1}+\int_{\mathcal{D}_{4-3}}dydz\int_{\alpha_-}^{\alpha_+} dx\mathbb{C}_{m,1}'\right\}I_{m,1}^\mathrm{div,m}\nonumber\\
&=&\frac{\alpha G_F^2|V_{us}|^2 M_K^5}{64\pi^4}\ln\frac{m_e^2}{\mu^2}\int_{2\sqrt{r_\pi}}^{1+r_\pi}dz f_+^2(t)\left\{\int_{c(z)-d(z)}^{c(z)+d(z)}dy\int_0^{\alpha_+}dx\bar{\mathbb{C}}(x,y,z)\right.\nonumber\\
&&\left.+\int_{0}^{c(z)-d(z)}dy\int_{\alpha_-}^{\alpha_+}dx\bar{\mathbb{C}}'(x,y,z)\right\}~.
\end{eqnarray} and the sum of divergent contributions to the total $K_{e3}$ decay rate becomes:
\begin{eqnarray}
(\delta\Gamma)^\mathrm{div,m}&=&(\delta\Gamma)_{(\delta f_+)^\mathrm{div,m}}+(\delta\Gamma)_{I_i^\mathrm{div,m}}+(\delta \Gamma)_{I_{m,1}^\mathrm{div,m}}\nonumber\\
&=&\frac{\alpha G_F^2|V_{us}|^2 M_K^5}{64\pi^4}\ln\frac{m_e^2}{\mu^2}\int_{2\sqrt{r_\pi}}^{1+r_\pi}dzf_+^2(t)\biggl\{\int_{c(z)-d(z)}^{c(z)+d(z)}dy\left[\frac{3}{2}+\ln\frac{\alpha_+^2}{(1-z+r_\pi)^2}\right]\nonumber\\
&&\times[r_\pi+(y-1)(y+z-1)]+\int_{c(z)-d(z)}^{c(z)+d(z)}dy\int_0^{\alpha_+}dx\bar{\mathbb{C}}(x,y,z)\nonumber\\
&&+\int_{0}^{c(z)-d(z)}dy\int_{\alpha_-}^{\alpha_+}dx\bar{\mathbb{C}}'(x,y,z)\biggr\}~.
\end{eqnarray}
The sum would vanish if the terms in the curly brackets add up to zero for all values of $z$. Indeed that happens, by direct integration one can show that:
\begin{eqnarray}
&&\int_{c(z)-d(z)}^{c(z)+d(z)}dy\left[\frac{3}{2}+\ln\frac{\alpha_+^2}{(1-z+r_\pi)^2}\right][r_\pi+(y-1)(y+z-1)]\nonumber\\
&=&\frac{1}{36}(z^2-4r_\pi)^{3/2}\left[12\ln\left(\frac{2-z+\sqrt{z^2-4r_\pi}}{2\sqrt{z^2-4r_\pi}}\right)+1\right]\equiv g_1(z)~,\nonumber\\
&&\int_{c(z)-d(z)}^{c(z)+d(z)}dy\int_0^\mathrm{\alpha_+}dx\bar{\mathbb{C}}(x,y,z)\nonumber\\
&=&\frac{1}{36}\biggl\{-4r_\pi\left[2\sqrt{z^2-4r_\pi}+9z-18\right]+8(z^2-3z+3)\sqrt{z^2-4r_\pi}+9z^2(z-2)\nonumber\\
&&+6\left[(z^2-6z+2r_\pi+6)\sqrt{z^2-4r_\pi}+(z-2)^3\right]\ln\left(\frac{2\sqrt{1-z+r_\pi}}{2-z-\sqrt{z^2-4r_\pi}}\right)\biggr\}\equiv g_2(z)~,\nonumber\\
\end{eqnarray}
and 
\begin{equation}
\int_{0}^{c(z)-d(z)}dy\int_{\alpha_-}^\mathrm{\alpha_+}dx\bar{\mathbb{C}}'(x,y,z)=-g_1(z)-g_2(z)~.
\end{equation}
So, we have $(\delta \Gamma)^\mathrm{div,m}=0$. This calculation shows that the full numerical result of the $K_{e3}$ decay rate in Refs.~\cite{Seng:2021boy,Seng:2021wcf,Seng:2022wcw} is free from photonic IR divergences and electron mass singularities. It supports the small error estimate given there for the 1 loop RC by demonstrating that no potentially numerically-large contributions enhanced by $\ln m_e$ were missed.

\section{\label{sec:extra}Extra finite terms when $\boldsymbol{m_e}$ is infinitesimal but non-zero}

In the previous section the emphasis was on terms with $\ln M_\gamma$ divergences and/or $\ln m_e$ singularities, assuming that all the remaining terms without such divergences would not distinguish between the mass-expansion and DR in the zero mass limit. However, the way equality comes about is quite novel and
particularly interesting. We will show in this section that there exists a contribution from the bremsstrahlung that remains finite when $m_e\rightarrow 0$, i.e. exists in the mass-expansion method, but naively appears to be absent when we set $m_e=0$ from the beginning. Its presence in DR will be subsequently discussed.

So far, in the bremsstrahlung process, we only examined the integrals $I_{m,1}$ that give a $\ln m_e$-divergence. However, there exists another set of integrals $I_{m,2}$ that gives a potentially more severe, $1/m_e^2$-power divergence: 
\begin{eqnarray}
I_{m,2}(p_i,p_e)&=&8\pi\int d\Gamma_k d\Gamma_{p_\nu}\frac{(2\pi)^4\delta^{(4)}(P-k-p_\nu)}{(p_i\cdot k)^m(p_e\cdot k)^2}\nonumber\\
&=&\frac{4(2p_e\cdot P)^m}{m_e^2(M_K^2x)^{m+1}(p_i\cdot p_e)^m}+\mathcal{O}(m_e^0)~.
\end{eqnarray}
Upon inspection, one finds that their coefficients $\mathbb{C}^{(\prime)}_{m,2}(x,y,z)$ always contain a factor $m_e^2$, i.e. they are chirally-suppressed in the squared-amplitude level.  Therefore, in the $m_e\rightarrow 0$ limit such terms make a finite contribution to the total decay rate. In other words, they represent a novel non-vanishing chirality-breaking effect in the $m_e\rightarrow 0$ limit. 
We emphasize that such contributions exists only in the mass-expansion method and not in DR, because in the latter the coefficients $\mathbb{C}^{(\prime)}_{m,2}$ are identically zero as we have set $m_e=0$ from the beginning. The existence of such  phenomena were recognized in the literature, e.g. Refs.~\cite{Gorsky:1989qd,Smilga:1990uq}, where it was argued that the $m_e=0$ is not smooth for certain observables in quantum electrodynamics because of such terms. Our interpretation will be
somewhat different. However, we will first evaluate the magnitude
of those contributions to the $K_{e3}$ radiative inclusive decay rate.

To account for the effects of such terms, 
we may define a new set of coefficients $\bar{\mathbb{D}}^{(\prime)}(x,y,z)$ by:
\begin{equation}
\lim_{m_e\rightarrow 0}\sum_m\mathbb{C}_{m,2}^{(\prime)}(x,y,z)\left\{\frac{4(2p_e\cdot P)^m}{m_e^2(M_K^2 x)^{m+1}(p_i\cdot p_e)^m}\right\}=64\pi\alpha M_K^2G_F^2|V_{us}|^2f_+^2(t)\bar{\mathbb{D}}^{(\prime)}(x,y,z)~.
\end{equation} 
Again, these coefficients turn out to be channel-independent:
\begin{eqnarray}
\bar{\mathbb{D}}(x,y,z)&=&\frac{2y[r_\pi(2y+z-2)-x(y+z-2)-(z-1)(2y+z-2)]}{(1-x-z+r_\pi)^2},\nonumber\\
\bar{\mathbb{D}}^{(\prime)}(x,y,z)&=&\bar{\mathbb{D}}(x,y,z)+\frac{2[r_\pi+(y-1)(y+z-1)]}{x}~.
\end{eqnarray}
With them, the extra finite contribution to the decay rate in the mass-expansion method is given in the $m_e\rightarrow 0$ limit as:
\begin{eqnarray}
(\delta \Gamma)_{\mathrm{ext,m}}&=&\frac{M_K^3}{4096\pi^5}\lim_{m_e\rightarrow 0}\sum_m\left\{\int_{\mathcal{D}_3}dydz\int_0^{\alpha_+} dx\mathbb{C}_{m,2}+\int_{\mathcal{D}_{4-3}}dydz\int_{\alpha_-}^{\alpha_+} dx\mathbb{C}_{m,2}'\right\}I_{m,2}\nonumber\\
&=&\frac{\alpha G_F^2|V_{us}|^2M_K^5}{64\pi^4}\int_{2\sqrt{r_\pi}}^{1+r_\pi}dzf_+^2(t)\left\{\int_{c(z)-d(z)}^{c(z)+d(z)}dy\int_0^{\alpha_+}dx\bar{\mathbb{D}}(x,y,z)\right.\nonumber\\
&&\left.+\int_{0}^{c(z)-d(z)}dy\int_{\alpha_-}^{\alpha_+}dx\bar{\mathbb{D}}'(x,y,z)\right\}~.
\end{eqnarray} 
Direct integration of the terms in the curly brackets returns a rather elegant expression:
\begin{eqnarray}
&&\int_{c(z)-d(z)}^{c(z)+d(z)}dy\int_0^{\alpha_+}dx\bar{\mathbb{D}}(x,y,z)+\int_{0}^{c(z)-d(z)}dy\int_{\alpha_-}^{\alpha_+}dx\bar{\mathbb{D}}'(x,y,z)\nonumber\\
&=&\frac{1}{18}(z^2-4r_\pi)^{3/2}\left\{3\ln\left[\frac{(z^2-4r_\pi)(2-z-\sqrt{z^2-4r_\pi})}{(1-z+r_\pi)(2-z+\sqrt{z^2-4r_\pi})}\right]+1\right\}\equiv H(z)~.\label{eq:Hz}
\end{eqnarray}
Numerically, utilizing a simple monopole parameterization of $f_+(t)$~\cite{NA482:2018rgv} one obtains $(\delta\Gamma)_\mathrm{ext,m}/\Gamma_{K_{e3}}\approx -0.01$\%, making up a negligibly small part in the full long-distance EM correction $\delta_\mathrm{EM}^{K\ell}$, which is 1.16(3)\% in the $K^0e$ channel and 0.21(5)\% in the $K^+e$ channel respectively. In Refs.~\cite{Seng:2021boy,Seng:2021wcf,Seng:2022wcw}, this contribution was automatically included in the numerical results because no $m_e=0$ simplifications were made in the analytic expressions.   

\section{\label{sec:DR}The same calculation with dimensional regularization}

Now we proceed to the second goal of this paper, namely to demonstrate the full equivalence between DR and the mass-expansion in reproducing the finite terms of the total radiatively inclusive $K_{e3}$ decay rate in the zero electron mass limit. 

We begin by setting $M_\gamma=m_e=0$ from the beginning, and generalize the space-time dimension to $d=4-\epsilon$. Some obvious generalizations of the basic formula in the previous section are:
\begin{equation}
\frac{d^4q'}{(2\pi)^4}\rightarrow \mu^\epsilon \frac{d^dq'}{(2\pi)^d}\label{eq:loopreplace}
\end{equation} 
in the loop integrals, and
\begin{equation}
(2\pi)^4\delta^{(4)}(k)\rightarrow \mu^{-\epsilon}(2\pi)^{d}\delta^{(d)}(k)~,~\frac{d^3k}{(2\pi)^3 2E_k}\rightarrow \mu^\epsilon\frac{d^{d-1}k}{(2\pi)^{d-1}2E_k}\equiv \mu^\epsilon d\Gamma_k
\end{equation}
in the phase space integrals. This is, however, not the full story. We will now demonstrate that extra modifications of the phase space formula are required to reproduce the correct finite pieces.

\subsection{\label{sec:PSDR}Three- and four-body phase space in $\boldsymbol{d}$-dimension}

We start by considering the following Lorentz-invariant integral:
\begin{equation}
I=\frac{1}{2M_K}\mu^\epsilon\int d\Gamma_{p'}\mu^\epsilon\int d\Gamma_{p_e}A(x,y,z)
\end{equation}
where $A(x,y,z)$ is an arbitrary scalar function of $x,y,z$. This is a straightforward generalization of the integral defined in Eq.(A.1) of Ref.~\cite{Seng:2021wcf}.
In general, the integral measure of the $d-1$ spatial components of a momentum $\vec{k}$ can be written as~\cite{Marciano:1974tv}:
\begin{equation}
\int d^{d-1}k=\int d|\vec{k}||\vec{k}|^{d-2}\left(\prod_{a=1}^{d-3}\int_0^\pi d\theta_a\sin^{d-a-2}\theta_a\right)\int_0^{2\pi}d\theta_{d-2}~.
\end{equation}
Now, since $A(x,y,z)$ is Lorentz-invariant, we may evaluate it in the kaon rest-frame, namely:
\begin{equation}
\vec{p}=\vec{0},\:\:\vec{p}'=|\vec{p}'|(1,0,0,...)\,\:\:\vec{p}_e=E_e(\cos\theta_e,\sin\theta_e,0,...)~.\label{eq:restframe}
\end{equation}
In this way, we can integrate all angles in $\vec{p}'$ and all except the first angle in $\vec{p}_e$ using the formula:
\begin{equation}
\int_0^\pi d\theta \sin^m\theta=\sqrt{\pi}\frac{\Gamma\left(\frac{1}{2}(m+1)\right)}{\Gamma\left(\frac{1}{2}(m+2)\right)}~,
\end{equation}
which gives:
\begin{equation}
I=\frac{1}{2M_K}\frac{\sqrt{\pi}}{64\pi^4\Gamma\left(\frac{3-\epsilon}{2}\right)\Gamma\left(\frac{2-\epsilon}{2}\right)}\int dE'dE_e\int_{-1}^{+1}d\cos\theta_e\left(\frac{16\pi^2\mu^4}{|\vec{p}'|^2E_e^2(1-\cos^2\theta_e)}\right)^{\epsilon/2}E_e|\vec{p}'|A(x,y,z)~.
\end{equation}

To further proceed, we notice that Eq.~\eqref{eq:restframe} allows us to express $E'$, $E_e$ and $\cos\theta_e$ using $x,y,z$. In particular, we find:
\begin{equation}
|\vec{p}'|^2E_e^2(1-\cos^2\theta_e)=M_K^4f(x,y,z)~,
\end{equation}
where
\begin{equation}
f(x,y,z)\equiv \frac{1}{4}\left[-r_\pi^2+r_\pi(2x-y(y+z-2)+2(z-1))-(x+y+z-1)(x-yz+y+z-1)\right]~.
\end{equation} 
With this definition, we obtain:
\begin{equation}
I=K\frac{M_K^3}{512\pi^4}\int_{2\sqrt{r_\pi}}^\infty dz\int_0^\infty dy\int_{\alpha_-}^{\alpha_+}dx [f(x,y,z)]^{-\epsilon/2}A(x,y,z)~,
\end{equation}
where
\begin{equation}
K\equiv\frac{\sqrt{\pi}}{2\Gamma\left(\frac{3-\epsilon}{2}\right)\Gamma\left(\frac{2-\epsilon}{2}\right)}\left(\frac{16\pi^2\mu^4}{M_K^4}\right)^{\epsilon/2}
\end{equation} 
is an overall constant multiplicative factor, which equals 1 when $\epsilon=0$. Although the equation above is defined by choosing a specific frame, since the final form is explicitly Lorentz-invariant, it obviously holds in every frame. This expression is analogous to the one in Eq.~(A.1) of Ref.~\cite{Seng:2021wcf}, so we may follow exactly the same logic in Appendix~A of that paper to derive the three- and four-body phase space formula in $d$-dimension:
\begin{eqnarray}
\Gamma_{3-\mathrm{body}}&=&K\frac{M_K}{256\pi^3}\int_{\mathcal{D}_3}dydz(f(0,y,z))^{-\epsilon/2}|M|_{K\rightarrow\pi e\nu}^2\nonumber\\
\Gamma_{4-\mathrm{body}}&=&K\frac{M_K^3}{512\pi^4}\left\{\int_{\mathcal{D}_3}dydz\int_0^{\alpha_+}dx+\int_{\mathcal{D}_{4-3}}dydz\int_{\alpha_-}^{\alpha_+}dx\right\}\mu^\epsilon\int d\Gamma_k d\Gamma_{p_\nu}\nonumber\\
&&\times (2\pi)^d\delta^{(d)}(P-k-p_\nu)(f(x,y,z))^{-\epsilon/2}|M|_{K\rightarrow\pi e\nu\gamma}^2~.
\end{eqnarray}
There are two new ingredients: (1) the overall multiplicative factor $K$, and (2) the factor $(f(x,y,z))^{-\epsilon/2}$ in the integrand. The former is irrelevant as long as the final decay rate is singularity-free, but the latter 
is important in order to get the correct finite contributions.  

It turns out that if we keep $m_e$ finite (i.e. do not pay attention to mass singularities), then the two new ingredients above are irrelevant (which is implicitly assumed in Ref.~\cite{Seng:2021wcf}). The reasons are of twofold: (1) when there are only IR singularities, the poles in both virtual and real corrections are only at the order $(2/\epsilon)$, and (2) these poles already cancel each other upon the $x$-integration, without touching $y$ and $z$. To explain this idea more clearly, let us write:
\begin{eqnarray}
&&\Gamma_{3-\mathrm{body}}+\Gamma_{4-\mathrm{body}}\nonumber\\
&=&K\frac{M_K}{256\pi^3}\int_{\mathcal{D}_3}dydz(f(0,y,z))^{-\epsilon/2}\Bigl\{|M|^2_{K\rightarrow\pi e\nu}+\frac{M_K^2}{2\pi}\int_0^{\alpha_+}dx\mu^\epsilon\int d\Gamma_k d\Gamma_{p_\nu}\nonumber\\
&&\times(2\pi)^d\delta^{(d)}(P-k-p_\nu)|M|^2_{K\rightarrow\pi e\nu\gamma}\Bigr\}+K\frac{M_K^3}{512\pi^4}\int_{\mathcal{D}_3}dydz\int_0^{\alpha_+}dx\mu^\epsilon\int d\Gamma_k d\Gamma_{p_\nu}\nonumber\\
&&\times(2\pi)^d\delta^{(d)}(P-k-p_\nu)\left\{(f(x,y,z))^{-\epsilon/2}-(f(0,y,z))^{-\epsilon/2}\right\}|M|_{K\rightarrow\pi e\nu\gamma}^2\nonumber\\
&&+K\frac{M_K^3}{512\pi^4}\int_{\mathcal{D}_{4-3}}dydz\int_{\alpha_-}^{\alpha_+}dx\mu^\epsilon\int d\Gamma_k d\Gamma_{p_\nu}(2\pi)^d\delta^{(d)}(P-k-p_\nu)(f(x,y,z))^{-\epsilon/2}|M|_{K\rightarrow\pi e\nu\gamma}^2~.\nonumber\\
\end{eqnarray}
There are three terms on the right-hand side, and the IR divergence occurs only within the curly brackets of the first term; in particular, in bremsstrahlung it appears as a log-divergence in the $x$-integral at $x\rightarrow 0$. The claim is that the IR divergences within the two terms in the curly bracket cancel out each other, so we can set $\epsilon=0$ in the remaining parts of the first integral, which brings $K$, $(f(0,y,z))^{-\epsilon/2}$ both to 1. In the second term, the factor $(f(x,y,z))^{-\epsilon/2}-(f(0,y,z))^{-\epsilon/2}$ renders the $x$-integration finite, so we can set $\epsilon=0$ from the beginning, which simply kills the entire term. Finally, the third term is by itself IR-finite; so, we can again take $\epsilon=0$ which brings $K$, $(f(x,y,z))^{-\epsilon/2}$ to 1. This proves our assertion.

Things are more complicated when mass singularities are also regularized using DR, because now one obtains poles of the order $(2/\epsilon)^2$ and $(2/\epsilon)$; the former cancels within the $x$-integration, but the latter only cancels upon the $y$-integration, therefore $(f(x,y,z))^{-\epsilon/2}$ gives a non-zero finite contribution which is the analog of the small chiral breaking contribution found in the massive electron calculation, as we shall show. 

\subsection{Divergent quantities in virtual corrections}

We may study the total $K_{e3}$ decay rate at $m_e=M_\gamma=0$ in Sirlin's representation using DR, starting from the virtual corrections. As an effective bookkeeping method for possible finite differences with the mass-expansion method, we apply the following strategy: for every quantity $F$ with singularities, we always make the following separation:
\begin{equation}
F=F^\mathrm{div,DR}+F^\mathrm{fin}+\mathcal{O}(\epsilon)~,
\end{equation}
where $F^\mathrm{fin}$ is defined to be \textit{exactly the same as that in the mass-expansion method} (see Eq.~\eqref{eq:Fmass}), which means $F^{\mathrm{div,DR}}$ may contain finite terms that are not proportional to $(2/\epsilon)^2$ or $(2/\epsilon)$. In this way, the comparison between the finite terms in DR and mass-expansion method can simply be done by comparing the effect of $F^\mathrm{div,DR}$ and $F^\mathrm{div,m}$ in the total decay rate.

One more technical detail is in order. In Sirlin's representation, one utilizes the following identity:
\begin{equation}
\gamma^\mu\gamma^\nu\gamma^\alpha=g^{\mu\nu}\gamma^\alpha-g^{\mu\alpha}\gamma^\nu+g^{\nu\alpha}\gamma^\mu-i\epsilon^{\mu\nu\alpha\beta}\gamma_\beta\gamma_5\label{eq:3gamma}
\end{equation}
to split the full $\gamma W$-box diagram into $\delta M_{\gamma W}^a$ and $\delta M_{\gamma W}^b$; the former is combined with $\delta M_2$ such that the residual integrals are ultraviolet-insensitive. A conceptual problem then arises when standard dimensional regularization (which we shall abbreviate as ``DReg'' to be specific) is applied to the formalism, because in DReg the generalization into $d$-dimensional vectors applies both to momenta and to the Dirac matrices $\gamma^\mu$, and the latter renders the totally-antisymmetric tensor $\epsilon^{\mu\nu\alpha\beta}$ ill-defined. This is more of a problem since there is a mass singularity in $\delta M_{\gamma W}^b$ (which contains an antisymmetric tensor) that has to be canceled with that in $(\delta M_2+\delta M_{\gamma W}^a)_\mathrm{int}$, as we discussed in Sec.~\ref{sec:Born}. Although in transforming the $\delta M_{\gamma W}^b$ contribution into $\delta f_+$ one again makes use of Eq.~\eqref{eq:3gamma} to get rid of the antisymmetric tensor, so it is possible that the final result is unambiguous since we have made ``even number of mistakes'', but it is still important to keep this in mind in order to track down possible extra finite terms in the final result. 

An approach to bypass the ambiguity above is to adopt the ``dimensional reduction'' (which we abbreviate as ``DRed'' to be specific) formalism~\cite{Siegel:1979wq}, where only momenta are treated in $d$-dimension whereas the Dirac matrices remain 4-vectors. In this way the $\epsilon$-tensor can be rigorously defined, but the formalism itself has its own problems, e.g. mathematical inconsistencies~\cite{Siegel:1980qs}. To stay on the safe side, in this work we apply both methods in the analysis; the finite difference between the two methods comes mainly from the fact that $\gamma^\mu\gamma_\alpha \gamma_\mu$ equals $(2-d)\gamma_\alpha$ in DReg but $-2\gamma_\alpha$ in DRed. In what follows we label both methods collectively as ``DR'', but will specify ``DReg'' or ``DRed'' whenever they make a difference.

Now we present the essential results of our DR analysis to the virtual corrections, where the replacement in Eq.\eqref{eq:loopreplace} applies to all loop integrals. First, the divergent part of the electron wavefunction renormalization reads:
\begin{equation}
\delta Z_e^\mathrm{div,DR}=-\frac{\alpha}{4\pi}R\left[-\frac{2}{\epsilon}-5+\delta\right]~,
\end{equation}
with $\delta$ equals 1 in DReg and 0 in DRed. Here we have defined:
\begin{equation}
R\equiv \left(\frac{e^{\gamma_E}}{4\pi}\right)^{-\epsilon/2}
\end{equation}
as a convenient multiplicative constant to absorb the effect of $\gamma_E$ and $\ln 4\pi$ which always appear in DR. Next, for the analytic piece in $\delta M_2+\delta M_{\gamma W}^a$ we have:
\begin{equation}
(\delta M_2+\delta M_{\gamma W}^a)_\mathrm{ana}^\mathrm{div,DR}=-\frac{\alpha}{2\pi}R\left(\frac{2}{\epsilon}+2\right)M_0~.
\end{equation}
Next we study the sum of the two-point and three-point function; since it involves only IR divergence and no mass singularity, there is a universal matching relation between these two schemes~\cite{Marciano:1974tv}: $\ln(M_\gamma^2/\mu^2)\leftrightarrow R(2/\epsilon)$. Applying this to Eq.~\eqref{eq:2pt+3pt} gives:
\begin{equation}
(\delta f_+)_\mathrm{2pt+3pt}^\mathrm{div,DR}=-\frac{\alpha}{4\pi}R\left(\frac{2}{\epsilon}\right)f_+~.
\end{equation}
And finally, the divergent part of the scalar function $C_0$ reads:
\begin{equation}
C_0^\mathrm{div,DR}=\frac{R}{M_1^2-v}\left\{-\frac{1}{2}\left(\frac{2}{\epsilon}\right)^2+\left(\frac{2}{\epsilon}\right)\left[\frac{1}{2}\ln\frac{M_1^2}{\mu^2}-\ln\frac{M_1^2}{M_1^2-v}\right]-\frac{\pi^2}{24}\right\}~.
\end{equation}

With all the above, the total divergent contribution to $\delta f_+$ is given in DR by:
\begin{eqnarray}
(\delta f_+)^\mathrm{div,DR}&=&\left\{\frac{1}{2}\delta Z_e^\mathrm{div,DR}-\frac{\alpha}{2\pi}(v-M_i^2)C_0^\mathrm{div,DR}-\frac{\alpha}{4\pi}R\left(\frac{2}{\epsilon}\right)\right\}f_+\nonumber\\
&=&-\frac{\alpha}{4\pi}R\left\{\left(\frac{2}{\epsilon}\right)^2+\left(\frac{2}{\epsilon}\right)\left(\frac{5}{2}-\ln\frac{M_i^2}{\mu^2}+2\ln\frac{M_i^2}{M_i^2-v}\right)+\frac{3+\delta}{2}+\frac{\pi^2}{12}\right\}f_+~,\nonumber\\
\end{eqnarray}
which is to be compared with the same quantity in the mass-expansion method, Eq.\eqref{eq:deltafpmass}. One observes that, unlike the case of finite $m_e$, in the presence of collinear mass singularity there is no simple matching that connects $\ln(m_e^2/\mu^2)$ and $\ln(M_\gamma^2/\mu^2)$ to $(2/\epsilon)$.

\subsection{Divergent integrals in bremsstrahlung} 

Next we study the divergent integrals in the bremsstrahlung process, starting from the integral $I_i$. Since the original definition in Eq.~\eqref{eq:Ii} involves the integration over $x$, in DR it must include the factor $(f(x,y,z))^{-\epsilon/2}$ due to the generalized phase space formula discussed in Sec.~\ref{sec:PSDR}. It is defined as:
\begin{eqnarray}
I_i&\equiv& \int_0^{\alpha_+}dx(f(x,y,z))^{-\epsilon/2}\mu^\epsilon\int d\Gamma_k d\Gamma_{p_\nu}(2\pi)^d\delta^{(d)}(P-k-p_\nu)\left(\frac{p_i}{p_i\cdot k}-\frac{p_e}{p_e\cdot k}\right)^2
\end{eqnarray}
where the divergent piece reads:
\begin{eqnarray}
I_i^\mathrm{div,DR}&=&\frac{R}{4\pi M_K^2}(f(0,y,z))^{-\epsilon/2}\biggl\{-\left(\frac{2}{\epsilon}\right)^2-\left(\frac{2}{\epsilon}\right)\left[1+\ln\left(\frac{\mu^2P_0^2(0)(1-\cos\chi)^2}{M_K^4\alpha_+^2}\right)\right]\nonumber\\
&&+2\mathrm{Li}_2\left(\frac{\alpha_+}{\alpha_-}\right)+\frac{\pi^2}{4}-2\ln^2\left(\frac{\alpha_+}{1-z+r_\pi}\right)-2\ln\left(\frac{\alpha_+}{1-z+r_\pi}\right)\nonumber\\
&&-4\mathrm{Li}_2\left(\frac{\alpha_+}{1-z+r_\pi}\right)\biggr\}~.
\end{eqnarray}

The remaining bremsstrahlung integrals with mass singularities are those in Eq.~\eqref{eq:Im1mass}, which are defined in DR as:
\begin{equation}
I_{m,1}(p_i)\equiv 8\pi\mu^\epsilon\int d\Gamma_k d\Gamma_{p_\nu}\frac{(2\pi)^d\delta^{(d)}(P-k-p_\nu)}{(p_i\cdot k)^mp_e\cdot k}~.
\end{equation}
Their divergent piece is given by:
\begin{equation}
I_{m,1}^\mathrm{div,DR}(p_i)=-\frac{2(2p_e\cdot P)^{m-1}}{(M_K^2x)^m (p_i\cdot p_e)^m}R\left(\frac{2}{\epsilon}+\ln\frac{(1-x-z+r_\pi)^2}{x^2}\right)~.
\end{equation}

\subsection{Reconciling DR and the mass-expansion}

Now we may collect all the aforementioned divergent contributions in DR to the total decay rate. They are:
\begin{eqnarray}
(\delta \Gamma)_{(\delta f_+)^\mathrm{div,DR}}&=&K\frac{M_K}{256\pi^3}\int_{\mathcal{D}_3}dydz(f(0,y,z))^{-\epsilon/2}\left\{\frac{2\mathfrak{Re}(\delta f_+)^\mathrm{div,DR}}{f_+}|M_0|^2(0,y,z)\right\},\nonumber\\
(\delta \Gamma)_{I_i^\mathrm{div,DR}}&=&K\frac{M_K^3}{512\pi^4}\int_{\mathcal{D}_3}dydz(-e^2)I_i^\mathrm{div,DR}|M_0|^2(0,y,z)~,\nonumber\\
(\delta \Gamma)_{I_{m,1}^\mathrm{div,DR}}&=&K\frac{M_K^3}{4096\pi^5}\sum_m\left\{\int_{\mathcal{D}_3}dydz\int_0^{\alpha_+} dx\mathbb{C}_{m,1}+\int_{\mathcal{D}_{4-3}}dydz\int_{\alpha_-}^{\alpha_+} dx\mathbb{C}_{m,1}'\right\}\nonumber\\
&&\times(f(x,y,z))^{-\epsilon/2}I_{m,1}^\mathrm{div,DR}~.
\end{eqnarray}
Based on the analysis in Sec.~\ref{sec:cancel}, it is easy to check that all the divergences of order $(2/\epsilon)^2$ and $(2/\epsilon)$ cancel upon summing the above three terms. and what left over is the extra finite terms in the DR method relative to the mass-expansion method. It is given in the $\epsilon\rightarrow 0$ limit by:
\begin{equation}
(\delta\Gamma)_\mathrm{ext,DR}=(\delta \Gamma)_{(\delta f_+)^\mathrm{div,DR}}+(\delta \Gamma)_{I_i^\mathrm{div,DR}}+(\delta \Gamma)_{I_{m,1}^\mathrm{div,DR}}
=\frac{\alpha G_F^2|V_{us}|^2M_K^5}{64\pi^4}\int_{2\sqrt{r_\pi}}^{1+r_\pi}dzf_+^2(t)h(z)~,
\end{equation}
where
\begin{eqnarray}
h(z)&\equiv&\int_{c(z)-d(z)}^{c(z)+d(z)}dy[r_\pi+(y-1)(y+z-1)]\Bigl[-\ln f(0,y,z)\left(\frac{3}{2}+\ln\frac{\alpha_+^2}{(1-z+r_\pi)^2}\right)\nonumber\\
&&+\frac{3+\delta}{2}+2\mathrm{Li}_2\left(\frac{\alpha_+}{\alpha_-}\right)+\frac{\pi^2}{3}-2\ln^2\left(\frac{\alpha_+}{1-z+r_\pi}\right)-2\ln\left(\frac{\alpha_+}{1-z+r_\pi}\right)\nonumber\\
&&-4\mathrm{Li}_2\left(\frac{\alpha_+}{1-z+r_\pi}\right)\Bigr]+\int_{c(z)-d(z)}^{c(z)+d(z)}dy\int_0^{\alpha_+}dx\Bigl[\bar{\mathbb{C}}(x,y,z)\Bigl(-\ln f(x,y,z)\nonumber\\
&&+\ln\frac{(1-x-z+r_\pi)^2}{x^2}\Bigr)+\delta\frac{4x f(x,y,z)}{(1-x-z+r_\pi)^3}\Bigr]+\int_0^{c(z)-d(z)}dy\int_{\alpha_-}^{\alpha_+}dx\nonumber\\
&&\times\Bigl[\bar{\mathbb{C}}'(x,y,z)\Bigl(-\ln f(x,y,z)+\ln\frac{(1-x-z+r_\pi)^2}{x^2}\Bigr)+\delta\frac{4x f(x,y,z)}{(1-x-z+r_\pi)^3}\Bigr]~.
\end{eqnarray}

Let us study the complicated expression above. First we observe that, apart from the electron wavefunction renormalization, there are also contributions from  bremsstrahlung in the $\mathcal{D}_3$ and $\mathcal{D}_{4-3}$ regions that are proportional to $\delta$, and thus differentiate DReg from DRed. This comes from a part of $|M_A|^2$:
\begin{equation}
|M_A|^2=-\frac{e^2G_F^2}{4}\left(\frac{1}{p_e\cdot k}\right)^2\mathrm{Tr}\left[(\slashed{P}-\slashed{k})\gamma^\mu\slashed{k}\gamma^\alpha\slashed{p}_e\gamma_\alpha\slashed{k}\gamma^\nu(1-\gamma_5)\right]F_\mu^{} F_\nu^*+....
\end{equation}
The two schemes give a difference in the value of $\gamma^\alpha\slashed{p}_e\gamma_\alpha$ at the order $(\epsilon/2)$, which then combines with the $(2/\epsilon)$-divergence in the $I_{m,1}$ integrals to give a finite difference. However, it is easy to check that all the terms proportional to $\delta$ add up to zero: 
\begin{eqnarray}
-\frac{1}{12}(z^2-4r_\pi)^{3/2}&=&\int_{c(z)-d(z)}^{c(z)+d(z)}dy\frac{1}{2}[r_\pi+(y-1)(y+z-1)]
\nonumber\\
&=&-\left(\int_{c(z)-d(z)}^{c(z)+d(z)}dy\int_0^{\alpha_+}dx+\int_0^{c(z)-d(z)}dy\int_{\alpha_-}^{\alpha_+}dx\right)\frac{4x f(x,y,z)}{(1-x-z+r_\pi)^3}~,\nonumber\\
\end{eqnarray}
which means DReg and DRed give the same total decay rate. This is an important confirmation of the formalism-independence of our result.

Finally, all the integrals in $h(z)$ can in fact be explicitly performed. The outcome turns out to be exactly equal to $H(z)$ in Eq.~\eqref{eq:Hz}:
\begin{equation}
h(z)=\frac{1}{18}(z^2-4r_\pi)^{3/2}\left\{3\ln\left[\frac{(z^2-4r_\pi)(2-z-\sqrt{z^2-4r_\pi})}{(1-z+r_\pi)(2-z+\sqrt{z^2-4r_\pi})}\right]+1\right\}=H(z)~,
\end{equation}
That equality can also be checked numerically. This implies $(\delta \Gamma)_\mathrm{ext,m}=(\delta \Gamma)_\mathrm{ext,DR}$, i.e. the DR method at $\epsilon\rightarrow 0$ agrees perfectly with the mass-expansion method at $m_e\rightarrow 0$ in the total, IR-finite $K_{e3}$ decay rate.  The effect of small chiral breaking terms
proportional to ${m_e^2}$ in the bremsstrahlung squared amplitude integral from collinear electron-photon configurations are reproduced by the space-time DR dependence.

\section{\label{sec:final}Final discussions}

In our discussion of the QED RC to ${K_{e3}}$ decays, we have tried to further clarify some aspects of the relatively recent improvements in the theory~\cite{Seng:2021boy,Seng:2021wcf,Seng:2022wcw} and to provide
cross-checks for the very small uncertainty, ${O(10^{-4})}$, they found. By introducing an electron mass expansion, we were able to demonstrate the cancellation of electron mass singularities in the radiative inclusive decay rates and verify
agreement with the KLN theorem. As a further check on the results,
a study of the inclusive rates for the case $m_e=0$, that is massless QED, was carried out using dimensional regularization of infrared photonic divergences and electron mass singularities.  In the $d\rightarrow 4$
limit, it confirmed the $m_e\rightarrow 0$ electron mass expansion results.
Particularly interesting is a very small contribution of about $-0.01\%$
from a set of $m_e^2$-suppressed terms in the bremsstrahlung squared amplitude that was subsequently power-enhanced by 1/${m_e^2}$ from the phase-space integration. That contribution was found to be needed for full agreement with the extra finite terms in the DR massless QED calculation that originate from the loop/bremsstrahlung integrals and from the non-trivial modification of the phase space formula in $d$ dimension. 

The major sources of theory uncertainty in the $K_{e3}$ RC are: (1) lattice QCD uncertainties in the mesonic $\gamma W$-box diagram that are quantifiable through standard procedures, and (2) residual hadronic structure-dependent uncertainties incalculable within Sirlin's representation (and not constrained by lattice). In particular, (2) is kept under control through a careful separation of the non-perturbative pieces from the perturbative ones. Our results in this work support the error analysis in Refs.~\cite{Seng:2021boy,Seng:2021wcf,Seng:2022wcw}, that all the $\ln m_e$-enhanced terms are fully contained in the non-perturbative pieces that were precisely calculated either analytically or numerically. The remaining, perturbative terms are free from $\ln m_e$-enhancement, hence the standard chiral power counting argument adopted in those papers provides a reliable estimation of its theory error.  
 
Given the smallness of the RC QED theory uncertainty, the emphasis should now be placed on other SM theory inputs. In particular, we strongly encourage more precise lattice QCD calculations of the form factor $f_+(t)$, not only its $t=0$ value but also its $t$-dependence. The latter can provide useful independent constraints on the $K\pi$ phase space factors. In addition, a lattice QCD calculation of $f_+(t)$ that is free of QED effects
avoids any issue of double counting which is potentially present in experimental measurements which presumably use an approximate Monte Carlo program to subtract out QED effects. Of course improved experimental measurements of all the kaon lifetimes and measurements of the radiative inclusive $K_{\ell 3}$ branching ratios, including the least-precise $K_S\mu$ channel, are also very desirable.

\begin{acknowledgments}

This work is supported in
part by the Deutsche Forschungsgemeinschaft (DFG, German Research
Foundation) and the NSFC through the funds provided to the Sino-German Collaborative Research Center TRR110 ``Symmetries and the Emergence of Structure in QCD'' (DFG Project-ID 196253076 - TRR 110, NSFC Grant No. 12070131001) (U-G.M and C.Y.S), by the Chinese Academy of Sciences (CAS) through a President's
International Fellowship Initiative (PIFI) (Grant No. 2018DM0034), by the VolkswagenStiftung
(Grant No. 93562) (U-G.M), and by the
U.S. Department of Energy under Grant DE-SC0012704 (WJM). 
  
\end{acknowledgments}

\begin{appendix}

\end{appendix}

\bibliography{Ke3_ref}

\begin{thebibliography}{91}
\expandafter\ifx\csname natexlab\endcsname\relax\def\natexlab#1{#1}\fi
\expandafter\ifx\csname bibnamefont\endcsname\relax
  \def\bibnamefont#1{#1}\fi
\expandafter\ifx\csname bibfnamefont\endcsname\relax
  \def\bibfnamefont#1{#1}\fi
\expandafter\ifx\csname citenamefont\endcsname\relax
  \def\citenamefont#1{#1}\fi
\expandafter\ifx\csname url\endcsname\relax
  \def\url#1{\texttt{#1}}\fi
\expandafter\ifx\csname urlprefix\endcsname\relax\def\urlprefix{URL }\fi
\providecommand{\bibinfo}[2]{#2}
\providecommand{\eprint}[2][]{\url{#2}}

\bibitem[{\citenamefont{Cabibbo}(1963)}]{Cabibbo:1963yz}
\bibinfo{author}{\bibfnamefont{N.}~\bibnamefont{Cabibbo}},
  \bibinfo{journal}{Phys. Rev. Lett.} \textbf{\bibinfo{volume}{10}},
  \bibinfo{pages}{531} (\bibinfo{year}{1963}).

\bibitem[{\citenamefont{Kobayashi and Maskawa}(1973)}]{Kobayashi:1973fv}
\bibinfo{author}{\bibfnamefont{M.}~\bibnamefont{Kobayashi}} \bibnamefont{and}
  \bibinfo{author}{\bibfnamefont{T.}~\bibnamefont{Maskawa}},
  \bibinfo{journal}{Prog. Theor. Phys.} \textbf{\bibinfo{volume}{49}},
  \bibinfo{pages}{652} (\bibinfo{year}{1973}).

\bibitem[{\citenamefont{Seng et~al.}(2022{\natexlab{a}})\citenamefont{Seng,
  Galviz, Gorchtein, and Mei\ss{}ner}}]{Seng:2022wcw}
\bibinfo{author}{\bibfnamefont{C.-Y.} \bibnamefont{Seng}},
  \bibinfo{author}{\bibfnamefont{D.}~\bibnamefont{Galviz}},
  \bibinfo{author}{\bibfnamefont{M.}~\bibnamefont{Gorchtein}},
  \bibnamefont{and} \bibinfo{author}{\bibfnamefont{U.-G.}
  \bibnamefont{Mei\ss{}ner}} (\bibinfo{year}{2022}{\natexlab{a}}),
  \eprint{2203.05217}.

\bibitem[{\citenamefont{Seng et~al.}(2018)\citenamefont{Seng, Gorchtein, Patel,
  and Ramsey-Musolf}}]{Seng:2018yzq}
\bibinfo{author}{\bibfnamefont{C.-Y.} \bibnamefont{Seng}},
  \bibinfo{author}{\bibfnamefont{M.}~\bibnamefont{Gorchtein}},
  \bibinfo{author}{\bibfnamefont{H.~H.} \bibnamefont{Patel}}, \bibnamefont{and}
  \bibinfo{author}{\bibfnamefont{M.~J.} \bibnamefont{Ramsey-Musolf}},
  \bibinfo{journal}{Phys. Rev. Lett.} \textbf{\bibinfo{volume}{121}},
  \bibinfo{pages}{241804} (\bibinfo{year}{2018}), \eprint{1807.10197}.

\bibitem[{\citenamefont{Seng et~al.}(2019)\citenamefont{Seng, Gorchtein, and
  Ramsey-Musolf}}]{Seng:2018qru}
\bibinfo{author}{\bibfnamefont{C.~Y.} \bibnamefont{Seng}},
  \bibinfo{author}{\bibfnamefont{M.}~\bibnamefont{Gorchtein}},
  \bibnamefont{and} \bibinfo{author}{\bibfnamefont{M.~J.}
  \bibnamefont{Ramsey-Musolf}}, \bibinfo{journal}{Phys. Rev.}
  \textbf{\bibinfo{volume}{D100}}, \bibinfo{pages}{013001}
  (\bibinfo{year}{2019}), \eprint{1812.03352}.

\bibitem[{\citenamefont{Czarnecki et~al.}(2019)\citenamefont{Czarnecki,
  Marciano, and Sirlin}}]{Czarnecki:2019mwq}
\bibinfo{author}{\bibfnamefont{A.}~\bibnamefont{Czarnecki}},
  \bibinfo{author}{\bibfnamefont{W.~J.} \bibnamefont{Marciano}},
  \bibnamefont{and} \bibinfo{author}{\bibfnamefont{A.}~\bibnamefont{Sirlin}},
  \bibinfo{journal}{Phys. Rev. D} \textbf{\bibinfo{volume}{100}},
  \bibinfo{pages}{073008} (\bibinfo{year}{2019}), \eprint{1907.06737}.

\bibitem[{\citenamefont{Gorchtein}(2019)}]{Gorchtein:2018fxl}
\bibinfo{author}{\bibfnamefont{M.}~\bibnamefont{Gorchtein}},
  \bibinfo{journal}{Phys. Rev. Lett.} \textbf{\bibinfo{volume}{123}},
  \bibinfo{pages}{042503} (\bibinfo{year}{2019}), \eprint{1812.04229}.

\bibitem[{\citenamefont{Seng et~al.}(2020{\natexlab{a}})\citenamefont{Seng,
  Feng, Gorchtein, and Jin}}]{Seng:2020wjq}
\bibinfo{author}{\bibfnamefont{C.-Y.} \bibnamefont{Seng}},
  \bibinfo{author}{\bibfnamefont{X.}~\bibnamefont{Feng}},
  \bibinfo{author}{\bibfnamefont{M.}~\bibnamefont{Gorchtein}},
  \bibnamefont{and} \bibinfo{author}{\bibfnamefont{L.-C.} \bibnamefont{Jin}},
  \bibinfo{journal}{Phys. Rev. D} \textbf{\bibinfo{volume}{101}},
  \bibinfo{pages}{111301} (\bibinfo{year}{2020}{\natexlab{a}}),
  \eprint{2003.11264}.

\bibitem[{\citenamefont{Shiells et~al.}(2021)\citenamefont{Shiells, Blunden,
  and Melnitchouk}}]{Shiells:2020fqp}
\bibinfo{author}{\bibfnamefont{K.}~\bibnamefont{Shiells}},
  \bibinfo{author}{\bibfnamefont{P.~G.} \bibnamefont{Blunden}},
  \bibnamefont{and}
  \bibinfo{author}{\bibfnamefont{W.}~\bibnamefont{Melnitchouk}},
  \bibinfo{journal}{Phys. Rev. D} \textbf{\bibinfo{volume}{104}},
  \bibinfo{pages}{033003} (\bibinfo{year}{2021}), \eprint{2012.01580}.

\bibitem[{\citenamefont{Hayen}(2021)}]{Hayen:2020cxh}
\bibinfo{author}{\bibfnamefont{L.}~\bibnamefont{Hayen}},
  \bibinfo{journal}{Phys. Rev. D} \textbf{\bibinfo{volume}{103}},
  \bibinfo{pages}{113001} (\bibinfo{year}{2021}), \eprint{2010.07262}.

\bibitem[{\citenamefont{Hardy and Towner}(2020)}]{Hardy:2020qwl}
\bibinfo{author}{\bibfnamefont{J.~C.} \bibnamefont{Hardy}} \bibnamefont{and}
  \bibinfo{author}{\bibfnamefont{I.~S.} \bibnamefont{Towner}},
  \bibinfo{journal}{Phys. Rev. C} \textbf{\bibinfo{volume}{102}},
  \bibinfo{pages}{045501} (\bibinfo{year}{2020}).

\bibitem[{\citenamefont{Seng et~al.}(2022{\natexlab{b}})\citenamefont{Seng,
  Galviz, Marciano, and Mei\ss{}ner}}]{Seng:2021nar}
\bibinfo{author}{\bibfnamefont{C.-Y.} \bibnamefont{Seng}},
  \bibinfo{author}{\bibfnamefont{D.}~\bibnamefont{Galviz}},
  \bibinfo{author}{\bibfnamefont{W.~J.} \bibnamefont{Marciano}},
  \bibnamefont{and} \bibinfo{author}{\bibfnamefont{U.-G.}
  \bibnamefont{Mei\ss{}ner}}, \bibinfo{journal}{Phys. Rev. D}
  \textbf{\bibinfo{volume}{105}}, \bibinfo{pages}{013005}
  (\bibinfo{year}{2022}{\natexlab{b}}), \eprint{2107.14708}.

\bibitem[{\citenamefont{Gonzalez-Alonso
  et~al.}(2019)\citenamefont{Gonzalez-Alonso, Naviliat-Cuncic, and
  Severijns}}]{Gonzalez-Alonso:2018omy}
\bibinfo{author}{\bibfnamefont{M.}~\bibnamefont{Gonzalez-Alonso}},
  \bibinfo{author}{\bibfnamefont{O.}~\bibnamefont{Naviliat-Cuncic}},
  \bibnamefont{and}
  \bibinfo{author}{\bibfnamefont{N.}~\bibnamefont{Severijns}},
  \bibinfo{journal}{Prog. Part. Nucl. Phys.} \textbf{\bibinfo{volume}{104}},
  \bibinfo{pages}{165} (\bibinfo{year}{2019}), \eprint{1803.08732}.

\bibitem[{\citenamefont{Bryman and
  Shrock}(2019{\natexlab{a}})}]{Bryman:2019ssi}
\bibinfo{author}{\bibfnamefont{D.}~\bibnamefont{Bryman}} \bibnamefont{and}
  \bibinfo{author}{\bibfnamefont{R.}~\bibnamefont{Shrock}},
  \bibinfo{journal}{Phys. Rev. D} \textbf{\bibinfo{volume}{100}},
  \bibinfo{pages}{053006} (\bibinfo{year}{2019}{\natexlab{a}}),
  \eprint{1904.06787}.

\bibitem[{\citenamefont{Bryman and
  Shrock}(2019{\natexlab{b}})}]{Bryman:2019bjg}
\bibinfo{author}{\bibfnamefont{D.}~\bibnamefont{Bryman}} \bibnamefont{and}
  \bibinfo{author}{\bibfnamefont{R.}~\bibnamefont{Shrock}},
  \bibinfo{journal}{Phys. Rev. D} \textbf{\bibinfo{volume}{100}},
  \bibinfo{pages}{073011} (\bibinfo{year}{2019}{\natexlab{b}}),
  \eprint{1909.11198}.

\bibitem[{\citenamefont{Belfatto et~al.}(2020)\citenamefont{Belfatto, Beradze,
  and Berezhiani}}]{Belfatto:2019swo}
\bibinfo{author}{\bibfnamefont{B.}~\bibnamefont{Belfatto}},
  \bibinfo{author}{\bibfnamefont{R.}~\bibnamefont{Beradze}}, \bibnamefont{and}
  \bibinfo{author}{\bibfnamefont{Z.}~\bibnamefont{Berezhiani}},
  \bibinfo{journal}{Eur. Phys. J. C} \textbf{\bibinfo{volume}{80}},
  \bibinfo{pages}{149} (\bibinfo{year}{2020}), \eprint{1906.02714}.

\bibitem[{\citenamefont{Tan}(2019)}]{Tan:2019yqp}
\bibinfo{author}{\bibfnamefont{W.}~\bibnamefont{Tan}} (\bibinfo{year}{2019}),
  \eprint{1906.10262}.

\bibitem[{\citenamefont{Grossman et~al.}(2020)\citenamefont{Grossman, Passemar,
  and Schacht}}]{Grossman:2019bzp}
\bibinfo{author}{\bibfnamefont{Y.}~\bibnamefont{Grossman}},
  \bibinfo{author}{\bibfnamefont{E.}~\bibnamefont{Passemar}}, \bibnamefont{and}
  \bibinfo{author}{\bibfnamefont{S.}~\bibnamefont{Schacht}},
  \bibinfo{journal}{JHEP} \textbf{\bibinfo{volume}{07}}, \bibinfo{pages}{068}
  (\bibinfo{year}{2020}), \eprint{1911.07821}.

\bibitem[{\citenamefont{Coutinho et~al.}(2020)\citenamefont{Coutinho,
  Crivellin, and Manzari}}]{Coutinho:2019aiy}
\bibinfo{author}{\bibfnamefont{A.~M.} \bibnamefont{Coutinho}},
  \bibinfo{author}{\bibfnamefont{A.}~\bibnamefont{Crivellin}},
  \bibnamefont{and} \bibinfo{author}{\bibfnamefont{C.~A.}
  \bibnamefont{Manzari}}, \bibinfo{journal}{Phys. Rev. Lett.}
  \textbf{\bibinfo{volume}{125}}, \bibinfo{pages}{071802}
  (\bibinfo{year}{2020}), \eprint{1912.08823}.

\bibitem[{\citenamefont{Falkowski et~al.}(2019)\citenamefont{Falkowski,
  Gonz\'alez-Alonso, and Tabrizi}}]{Falkowski:2019xoe}
\bibinfo{author}{\bibfnamefont{A.}~\bibnamefont{Falkowski}},
  \bibinfo{author}{\bibfnamefont{M.}~\bibnamefont{Gonz\'alez-Alonso}},
  \bibnamefont{and} \bibinfo{author}{\bibfnamefont{Z.}~\bibnamefont{Tabrizi}},
  \bibinfo{journal}{JHEP} \textbf{\bibinfo{volume}{05}}, \bibinfo{pages}{173}
  (\bibinfo{year}{2019}), \eprint{1901.04553}.

\bibitem[{\citenamefont{Cirigliano et~al.}(2019)\citenamefont{Cirigliano,
  Garcia, Gazit, Naviliat-Cuncic, Savard, and Young}}]{Cirgiliano:2019nyn}
\bibinfo{author}{\bibfnamefont{V.}~\bibnamefont{Cirigliano}},
  \bibinfo{author}{\bibfnamefont{A.}~\bibnamefont{Garcia}},
  \bibinfo{author}{\bibfnamefont{D.}~\bibnamefont{Gazit}},
  \bibinfo{author}{\bibfnamefont{O.}~\bibnamefont{Naviliat-Cuncic}},
  \bibinfo{author}{\bibfnamefont{G.}~\bibnamefont{Savard}}, \bibnamefont{and}
  \bibinfo{author}{\bibfnamefont{A.}~\bibnamefont{Young}}
  (\bibinfo{year}{2019}), \eprint{1907.02164}.

\bibitem[{\citenamefont{Jho et~al.}(2020)\citenamefont{Jho, Lee, Park, Park,
  and Tseng}}]{Jho:2020jsa}
\bibinfo{author}{\bibfnamefont{Y.}~\bibnamefont{Jho}},
  \bibinfo{author}{\bibfnamefont{S.~M.} \bibnamefont{Lee}},
  \bibinfo{author}{\bibfnamefont{S.~C.} \bibnamefont{Park}},
  \bibinfo{author}{\bibfnamefont{Y.}~\bibnamefont{Park}}, \bibnamefont{and}
  \bibinfo{author}{\bibfnamefont{P.-Y.} \bibnamefont{Tseng}},
  \bibinfo{journal}{JHEP} \textbf{\bibinfo{volume}{04}}, \bibinfo{pages}{086}
  (\bibinfo{year}{2020}), \eprint{2001.06572}.

\bibitem[{\citenamefont{Yue and Cheng}(2021)}]{Yue:2020wkj}
\bibinfo{author}{\bibfnamefont{C.~X.} \bibnamefont{Yue}} \bibnamefont{and}
  \bibinfo{author}{\bibfnamefont{X.~J.} \bibnamefont{Cheng}},
  \bibinfo{journal}{Nucl. Phys. B} \textbf{\bibinfo{volume}{963}},
  \bibinfo{pages}{115280} (\bibinfo{year}{2021}), \eprint{2008.10027}.

\bibitem[{\citenamefont{Endo and Mishima}(2020)}]{Endo:2020tkb}
\bibinfo{author}{\bibfnamefont{M.}~\bibnamefont{Endo}} \bibnamefont{and}
  \bibinfo{author}{\bibfnamefont{S.}~\bibnamefont{Mishima}},
  \bibinfo{journal}{JHEP} \textbf{\bibinfo{volume}{08}}, \bibinfo{pages}{004}
  (\bibinfo{year}{2020}), \eprint{2005.03933}.

\bibitem[{\citenamefont{Capdevila et~al.}(2021)\citenamefont{Capdevila,
  Crivellin, Manzari, and Montull}}]{Capdevila:2020rrl}
\bibinfo{author}{\bibfnamefont{B.}~\bibnamefont{Capdevila}},
  \bibinfo{author}{\bibfnamefont{A.}~\bibnamefont{Crivellin}},
  \bibinfo{author}{\bibfnamefont{C.~A.} \bibnamefont{Manzari}},
  \bibnamefont{and} \bibinfo{author}{\bibfnamefont{M.}~\bibnamefont{Montull}},
  \bibinfo{journal}{Phys. Rev. D} \textbf{\bibinfo{volume}{103}},
  \bibinfo{pages}{015032} (\bibinfo{year}{2021}), \eprint{2005.13542}.

\bibitem[{\citenamefont{Eberhardt et~al.}(2021)\citenamefont{Eberhardt,
  Mart\'\i{}nez, and Pich}}]{Eberhardt:2020dat}
\bibinfo{author}{\bibfnamefont{O.}~\bibnamefont{Eberhardt}},
  \bibinfo{author}{\bibfnamefont{A.~P.~n.} \bibnamefont{Mart\'\i{}nez}},
  \bibnamefont{and} \bibinfo{author}{\bibfnamefont{A.}~\bibnamefont{Pich}},
  \bibinfo{journal}{JHEP} \textbf{\bibinfo{volume}{05}}, \bibinfo{pages}{005}
  (\bibinfo{year}{2021}), \eprint{2012.09200}.

\bibitem[{\citenamefont{Cheung et~al.}(2020)\citenamefont{Cheung, Keung, Lu,
  and Tseng}}]{Cheung:2020vqm}
\bibinfo{author}{\bibfnamefont{K.}~\bibnamefont{Cheung}},
  \bibinfo{author}{\bibfnamefont{W.-Y.} \bibnamefont{Keung}},
  \bibinfo{author}{\bibfnamefont{C.-T.} \bibnamefont{Lu}}, \bibnamefont{and}
  \bibinfo{author}{\bibfnamefont{P.-Y.} \bibnamefont{Tseng}},
  \bibinfo{journal}{JHEP} \textbf{\bibinfo{volume}{05}}, \bibinfo{pages}{117}
  (\bibinfo{year}{2020}), \eprint{2001.02853}.

\bibitem[{\citenamefont{Crivellin et~al.}(2020)\citenamefont{Crivellin, Kirk,
  Manzari, and Montull}}]{Crivellin:2020ebi}
\bibinfo{author}{\bibfnamefont{A.}~\bibnamefont{Crivellin}},
  \bibinfo{author}{\bibfnamefont{F.}~\bibnamefont{Kirk}},
  \bibinfo{author}{\bibfnamefont{C.~A.} \bibnamefont{Manzari}},
  \bibnamefont{and} \bibinfo{author}{\bibfnamefont{M.}~\bibnamefont{Montull}},
  \bibinfo{journal}{JHEP} \textbf{\bibinfo{volume}{12}}, \bibinfo{pages}{166}
  (\bibinfo{year}{2020}), \eprint{2008.01113}.

\bibitem[{\citenamefont{Crivellin
  et~al.}(2021{\natexlab{a}})\citenamefont{Crivellin, Kirk, Manzari, and
  Panizzi}}]{Crivellin:2020klg}
\bibinfo{author}{\bibfnamefont{A.}~\bibnamefont{Crivellin}},
  \bibinfo{author}{\bibfnamefont{F.}~\bibnamefont{Kirk}},
  \bibinfo{author}{\bibfnamefont{C.~A.} \bibnamefont{Manzari}},
  \bibnamefont{and} \bibinfo{author}{\bibfnamefont{L.}~\bibnamefont{Panizzi}},
  \bibinfo{journal}{Phys. Rev. D} \textbf{\bibinfo{volume}{103}},
  \bibinfo{pages}{073002} (\bibinfo{year}{2021}{\natexlab{a}}),
  \eprint{2012.09845}.

\bibitem[{\citenamefont{Crivellin and Hoferichter}(2020)}]{Crivellin:2020lzu}
\bibinfo{author}{\bibfnamefont{A.}~\bibnamefont{Crivellin}} \bibnamefont{and}
  \bibinfo{author}{\bibfnamefont{M.}~\bibnamefont{Hoferichter}},
  \bibinfo{journal}{Phys. Rev. Lett.} \textbf{\bibinfo{volume}{125}},
  \bibinfo{pages}{111801} (\bibinfo{year}{2020}), \eprint{2002.07184}.

\bibitem[{\citenamefont{Kirk}(2021)}]{Kirk:2020wdk}
\bibinfo{author}{\bibfnamefont{M.}~\bibnamefont{Kirk}}, \bibinfo{journal}{Phys.
  Rev. D} \textbf{\bibinfo{volume}{103}}, \bibinfo{pages}{035004}
  (\bibinfo{year}{2021}), \eprint{2008.03261}.

\bibitem[{\citenamefont{Crivellin
  et~al.}(2021{\natexlab{b}})\citenamefont{Crivellin, Manzari, Alguero, and
  Matias}}]{Crivellin:2020oup}
\bibinfo{author}{\bibfnamefont{A.}~\bibnamefont{Crivellin}},
  \bibinfo{author}{\bibfnamefont{C.~A.} \bibnamefont{Manzari}},
  \bibinfo{author}{\bibfnamefont{M.}~\bibnamefont{Alguero}}, \bibnamefont{and}
  \bibinfo{author}{\bibfnamefont{J.}~\bibnamefont{Matias}},
  \bibinfo{journal}{Phys. Rev. Lett.} \textbf{\bibinfo{volume}{127}},
  \bibinfo{pages}{011801} (\bibinfo{year}{2021}{\natexlab{b}}),
  \eprint{2010.14504}.

\bibitem[{\citenamefont{Falkowski et~al.}(2021)\citenamefont{Falkowski,
  Gonz\'alez-Alonso, and Naviliat-Cuncic}}]{Falkowski:2020pma}
\bibinfo{author}{\bibfnamefont{A.}~\bibnamefont{Falkowski}},
  \bibinfo{author}{\bibfnamefont{M.}~\bibnamefont{Gonz\'alez-Alonso}},
  \bibnamefont{and}
  \bibinfo{author}{\bibfnamefont{O.}~\bibnamefont{Naviliat-Cuncic}},
  \bibinfo{journal}{JHEP} \textbf{\bibinfo{volume}{04}}, \bibinfo{pages}{126}
  (\bibinfo{year}{2021}), \eprint{2010.13797}.

\bibitem[{\citenamefont{Be\v{c}irevi\'c
  et~al.}(2021)\citenamefont{Be\v{c}irevi\'c, Jaffredo, Pe\~nuelas, and
  Sumensari}}]{Becirevic:2020rzi}
\bibinfo{author}{\bibfnamefont{D.}~\bibnamefont{Be\v{c}irevi\'c}},
  \bibinfo{author}{\bibfnamefont{F.}~\bibnamefont{Jaffredo}},
  \bibinfo{author}{\bibfnamefont{A.}~\bibnamefont{Pe\~nuelas}},
  \bibnamefont{and}
  \bibinfo{author}{\bibfnamefont{O.}~\bibnamefont{Sumensari}},
  \bibinfo{journal}{JHEP} \textbf{\bibinfo{volume}{05}}, \bibinfo{pages}{175}
  (\bibinfo{year}{2021}), \eprint{2012.09872}.

\bibitem[{\citenamefont{Crivellin
  et~al.}(2021{\natexlab{c}})\citenamefont{Crivellin, Hoferichter, and
  Manzari}}]{Crivellin:2021njn}
\bibinfo{author}{\bibfnamefont{A.}~\bibnamefont{Crivellin}},
  \bibinfo{author}{\bibfnamefont{M.}~\bibnamefont{Hoferichter}},
  \bibnamefont{and} \bibinfo{author}{\bibfnamefont{C.~A.}
  \bibnamefont{Manzari}}, \bibinfo{journal}{Phys. Rev. Lett.}
  \textbf{\bibinfo{volume}{127}}, \bibinfo{pages}{071801}
  (\bibinfo{year}{2021}{\natexlab{c}}), \eprint{2102.02825}.

\bibitem[{\citenamefont{Lubicz et~al.}(2009)\citenamefont{Lubicz, Mescia,
  Simula, and Tarantino}}]{Lubicz:2009ht}
\bibinfo{author}{\bibfnamefont{V.}~\bibnamefont{Lubicz}},
  \bibinfo{author}{\bibfnamefont{F.}~\bibnamefont{Mescia}},
  \bibinfo{author}{\bibfnamefont{S.}~\bibnamefont{Simula}}, \bibnamefont{and}
  \bibinfo{author}{\bibfnamefont{C.}~\bibnamefont{Tarantino}}
  (\bibinfo{collaboration}{ETM}), \bibinfo{journal}{Phys. Rev. D}
  \textbf{\bibinfo{volume}{80}}, \bibinfo{pages}{111502}
  (\bibinfo{year}{2009}), \eprint{0906.4728}.

\bibitem[{\citenamefont{Bazavov et~al.}(2013)}]{Bazavov:2012cd}
\bibinfo{author}{\bibfnamefont{A.}~\bibnamefont{Bazavov}} \bibnamefont{et~al.},
  \bibinfo{journal}{Phys. Rev. D} \textbf{\bibinfo{volume}{87}},
  \bibinfo{pages}{073012} (\bibinfo{year}{2013}), \eprint{1212.4993}.

\bibitem[{\citenamefont{Boyle et~al.}(2015)}]{Boyle:2015hfa}
\bibinfo{author}{\bibfnamefont{P.~A.} \bibnamefont{Boyle}} \bibnamefont{et~al.}
  (\bibinfo{collaboration}{RBC/UKQCD}), \bibinfo{journal}{JHEP}
  \textbf{\bibinfo{volume}{06}}, \bibinfo{pages}{164} (\bibinfo{year}{2015}),
  \eprint{1504.01692}.

\bibitem[{\citenamefont{Carrasco et~al.}(2016)\citenamefont{Carrasco, Lami,
  Lubicz, Riggio, Simula, and Tarantino}}]{Carrasco:2016kpy}
\bibinfo{author}{\bibfnamefont{N.}~\bibnamefont{Carrasco}},
  \bibinfo{author}{\bibfnamefont{P.}~\bibnamefont{Lami}},
  \bibinfo{author}{\bibfnamefont{V.}~\bibnamefont{Lubicz}},
  \bibinfo{author}{\bibfnamefont{L.}~\bibnamefont{Riggio}},
  \bibinfo{author}{\bibfnamefont{S.}~\bibnamefont{Simula}}, \bibnamefont{and}
  \bibinfo{author}{\bibfnamefont{C.}~\bibnamefont{Tarantino}},
  \bibinfo{journal}{Phys. Rev. D} \textbf{\bibinfo{volume}{93}},
  \bibinfo{pages}{114512} (\bibinfo{year}{2016}), \eprint{1602.04113}.

\bibitem[{\citenamefont{Bazavov et~al.}(2019)}]{Bazavov:2018kjg}
\bibinfo{author}{\bibfnamefont{A.}~\bibnamefont{Bazavov}} \bibnamefont{et~al.}
  (\bibinfo{collaboration}{Fermilab Lattice, MILC}), \bibinfo{journal}{Phys.
  Rev.} \textbf{\bibinfo{volume}{D99}}, \bibinfo{pages}{114509}
  (\bibinfo{year}{2019}), \eprint{1809.02827}.

\bibitem[{\citenamefont{Lichard}(1997)}]{Lichard:1997ya}
\bibinfo{author}{\bibfnamefont{P.}~\bibnamefont{Lichard}},
  \bibinfo{journal}{Phys. Rev. D} \textbf{\bibinfo{volume}{55}},
  \bibinfo{pages}{5385} (\bibinfo{year}{1997}), \eprint{hep-ph/9702345}.

\bibitem[{\citenamefont{Antonelli et~al.}(2010)}]{Antonelli:2010yf}
\bibinfo{author}{\bibfnamefont{M.}~\bibnamefont{Antonelli}}
  \bibnamefont{et~al.} (\bibinfo{collaboration}{FlaviaNet Working Group on Kaon
  Decays}), \bibinfo{journal}{Eur. Phys. J. C} \textbf{\bibinfo{volume}{69}},
  \bibinfo{pages}{399} (\bibinfo{year}{2010}), \eprint{1005.2323}.

\bibitem[{\citenamefont{Hill}(2006)}]{Hill:2006bq}
\bibinfo{author}{\bibfnamefont{R.~J.} \bibnamefont{Hill}},
  \bibinfo{journal}{Phys. Rev. D} \textbf{\bibinfo{volume}{74}},
  \bibinfo{pages}{096006} (\bibinfo{year}{2006}), \eprint{hep-ph/0607108}.

\bibitem[{\citenamefont{Bernard et~al.}(2006)\citenamefont{Bernard, Oertel,
  Passemar, and Stern}}]{Bernard:2006gy}
\bibinfo{author}{\bibfnamefont{V.}~\bibnamefont{Bernard}},
  \bibinfo{author}{\bibfnamefont{M.}~\bibnamefont{Oertel}},
  \bibinfo{author}{\bibfnamefont{E.}~\bibnamefont{Passemar}}, \bibnamefont{and}
  \bibinfo{author}{\bibfnamefont{J.}~\bibnamefont{Stern}},
  \bibinfo{journal}{Phys. Lett.} \textbf{\bibinfo{volume}{B638}},
  \bibinfo{pages}{480} (\bibinfo{year}{2006}), \eprint{hep-ph/0603202}.

\bibitem[{\citenamefont{Bernard et~al.}(2009)\citenamefont{Bernard, Oertel,
  Passemar, and Stern}}]{Bernard:2009zm}
\bibinfo{author}{\bibfnamefont{V.}~\bibnamefont{Bernard}},
  \bibinfo{author}{\bibfnamefont{M.}~\bibnamefont{Oertel}},
  \bibinfo{author}{\bibfnamefont{E.}~\bibnamefont{Passemar}}, \bibnamefont{and}
  \bibinfo{author}{\bibfnamefont{J.}~\bibnamefont{Stern}},
  \bibinfo{journal}{Phys. Rev.} \textbf{\bibinfo{volume}{D80}},
  \bibinfo{pages}{034034} (\bibinfo{year}{2009}), \eprint{0903.1654}.

\bibitem[{\citenamefont{Abouzaid et~al.}(2010)}]{Abouzaid:2009ry}
\bibinfo{author}{\bibfnamefont{E.}~\bibnamefont{Abouzaid}} \bibnamefont{et~al.}
  (\bibinfo{collaboration}{KTeV}), \bibinfo{journal}{Phys. Rev. D}
  \textbf{\bibinfo{volume}{81}}, \bibinfo{pages}{052001}
  (\bibinfo{year}{2010}), \eprint{0912.1291}.

\bibitem[{\citenamefont{Moulson}(2021)}]{PSCKM21}
\bibinfo{author}{\bibfnamefont{M.}~\bibnamefont{Moulson}}
  (\bibinfo{year}{2021}), \bibinfo{note}{$V_{us}$ from kaon decays,
  \textit{11th International Workshop on the CKM Unitarity Triangle (CKM
  2021)},
  \url{https://indico.cern.ch/event/891123/contributions/4601856/attachments/2351074/4011941/CKM
  202021.pdf}}.

\bibitem[{\citenamefont{Blum et~al.}(2016)}]{RBC:2014ntl}
\bibinfo{author}{\bibfnamefont{T.}~\bibnamefont{Blum}} \bibnamefont{et~al.}
  (\bibinfo{collaboration}{RBC, UKQCD}), \bibinfo{journal}{Phys. Rev. D}
  \textbf{\bibinfo{volume}{93}}, \bibinfo{pages}{074505}
  (\bibinfo{year}{2016}), \eprint{1411.7017}.

\bibitem[{\citenamefont{Durr et~al.}(2011{\natexlab{a}})\citenamefont{Durr,
  Fodor, Hoelbling, Katz, Krieg, Kurth, Lellouch, Lippert, Szabo, and
  Vulvert}}]{Durr:2010vn}
\bibinfo{author}{\bibfnamefont{S.}~\bibnamefont{Durr}},
  \bibinfo{author}{\bibfnamefont{Z.}~\bibnamefont{Fodor}},
  \bibinfo{author}{\bibfnamefont{C.}~\bibnamefont{Hoelbling}},
  \bibinfo{author}{\bibfnamefont{S.~D.} \bibnamefont{Katz}},
  \bibinfo{author}{\bibfnamefont{S.}~\bibnamefont{Krieg}},
  \bibinfo{author}{\bibfnamefont{T.}~\bibnamefont{Kurth}},
  \bibinfo{author}{\bibfnamefont{L.}~\bibnamefont{Lellouch}},
  \bibinfo{author}{\bibfnamefont{T.}~\bibnamefont{Lippert}},
  \bibinfo{author}{\bibfnamefont{K.~K.} \bibnamefont{Szabo}}, \bibnamefont{and}
  \bibinfo{author}{\bibfnamefont{G.}~\bibnamefont{Vulvert}},
  \bibinfo{journal}{Phys. Lett. B} \textbf{\bibinfo{volume}{701}},
  \bibinfo{pages}{265} (\bibinfo{year}{2011}{\natexlab{a}}),
  \eprint{1011.2403}.

\bibitem[{\citenamefont{Durr et~al.}(2011{\natexlab{b}})\citenamefont{Durr,
  Fodor, Hoelbling, Katz, Krieg, Kurth, Lellouch, Lippert, Szabo, and
  Vulvert}}]{Durr:2010aw}
\bibinfo{author}{\bibfnamefont{S.}~\bibnamefont{Durr}},
  \bibinfo{author}{\bibfnamefont{Z.}~\bibnamefont{Fodor}},
  \bibinfo{author}{\bibfnamefont{C.}~\bibnamefont{Hoelbling}},
  \bibinfo{author}{\bibfnamefont{S.~D.} \bibnamefont{Katz}},
  \bibinfo{author}{\bibfnamefont{S.}~\bibnamefont{Krieg}},
  \bibinfo{author}{\bibfnamefont{T.}~\bibnamefont{Kurth}},
  \bibinfo{author}{\bibfnamefont{L.}~\bibnamefont{Lellouch}},
  \bibinfo{author}{\bibfnamefont{T.}~\bibnamefont{Lippert}},
  \bibinfo{author}{\bibfnamefont{K.~K.} \bibnamefont{Szabo}}, \bibnamefont{and}
  \bibinfo{author}{\bibfnamefont{G.}~\bibnamefont{Vulvert}},
  \bibinfo{journal}{JHEP} \textbf{\bibinfo{volume}{08}}, \bibinfo{pages}{148}
  (\bibinfo{year}{2011}{\natexlab{b}}), \eprint{1011.2711}.

\bibitem[{\citenamefont{Bazavov et~al.}(2009)}]{MILC:2009ltw}
\bibinfo{author}{\bibfnamefont{A.}~\bibnamefont{Bazavov}} \bibnamefont{et~al.}
  (\bibinfo{collaboration}{MILC}), \bibinfo{journal}{PoS}
  \textbf{\bibinfo{volume}{CD09}}, \bibinfo{pages}{007} (\bibinfo{year}{2009}),
  \eprint{0910.2966}.

\bibitem[{\citenamefont{Fodor et~al.}(2016)\citenamefont{Fodor, Hoelbling,
  Krieg, Lellouch, Lippert, Portelli, Sastre, Szabo, and
  Varnhorst}}]{Fodor:2016bgu}
\bibinfo{author}{\bibfnamefont{Z.}~\bibnamefont{Fodor}},
  \bibinfo{author}{\bibfnamefont{C.}~\bibnamefont{Hoelbling}},
  \bibinfo{author}{\bibfnamefont{S.}~\bibnamefont{Krieg}},
  \bibinfo{author}{\bibfnamefont{L.}~\bibnamefont{Lellouch}},
  \bibinfo{author}{\bibfnamefont{T.}~\bibnamefont{Lippert}},
  \bibinfo{author}{\bibfnamefont{A.}~\bibnamefont{Portelli}},
  \bibinfo{author}{\bibfnamefont{A.}~\bibnamefont{Sastre}},
  \bibinfo{author}{\bibfnamefont{K.~K.} \bibnamefont{Szabo}}, \bibnamefont{and}
  \bibinfo{author}{\bibfnamefont{L.}~\bibnamefont{Varnhorst}},
  \bibinfo{journal}{Phys. Rev. Lett.} \textbf{\bibinfo{volume}{117}},
  \bibinfo{pages}{082001} (\bibinfo{year}{2016}), \eprint{1604.07112}.

\bibitem[{\citenamefont{Bazavov et~al.}(2018)}]{Bazavov:2017lyh}
\bibinfo{author}{\bibfnamefont{A.}~\bibnamefont{Bazavov}} \bibnamefont{et~al.},
  \bibinfo{journal}{Phys. Rev. D} \textbf{\bibinfo{volume}{98}},
  \bibinfo{pages}{074512} (\bibinfo{year}{2018}), \eprint{1712.09262}.

\bibitem[{\citenamefont{Carrasco et~al.}(2014)}]{EuropeanTwistedMass:2014osg}
\bibinfo{author}{\bibfnamefont{N.}~\bibnamefont{Carrasco}} \bibnamefont{et~al.}
  (\bibinfo{collaboration}{European Twisted Mass}), \bibinfo{journal}{Nucl.
  Phys. B} \textbf{\bibinfo{volume}{887}}, \bibinfo{pages}{19}
  (\bibinfo{year}{2014}), \eprint{1403.4504}.

\bibitem[{\citenamefont{Bazavov et~al.}(2014)}]{FermilabLattice:2014tsy}
\bibinfo{author}{\bibfnamefont{A.}~\bibnamefont{Bazavov}} \bibnamefont{et~al.}
  (\bibinfo{collaboration}{Fermilab Lattice, MILC}), \bibinfo{journal}{Phys.
  Rev. D} \textbf{\bibinfo{volume}{90}}, \bibinfo{pages}{074509}
  (\bibinfo{year}{2014}), \eprint{1407.3772}.

\bibitem[{\citenamefont{Giusti et~al.}(2017)\citenamefont{Giusti, Lubicz,
  Tarantino, Martinelli, Sanfilippo, Simula, and Tantalo}}]{Giusti:2017dmp}
\bibinfo{author}{\bibfnamefont{D.}~\bibnamefont{Giusti}},
  \bibinfo{author}{\bibfnamefont{V.}~\bibnamefont{Lubicz}},
  \bibinfo{author}{\bibfnamefont{C.}~\bibnamefont{Tarantino}},
  \bibinfo{author}{\bibfnamefont{G.}~\bibnamefont{Martinelli}},
  \bibinfo{author}{\bibfnamefont{F.}~\bibnamefont{Sanfilippo}},
  \bibinfo{author}{\bibfnamefont{S.}~\bibnamefont{Simula}}, \bibnamefont{and}
  \bibinfo{author}{\bibfnamefont{N.}~\bibnamefont{Tantalo}},
  \bibinfo{journal}{Phys. Rev. D} \textbf{\bibinfo{volume}{95}},
  \bibinfo{pages}{114504} (\bibinfo{year}{2017}), \eprint{1704.06561}.

\bibitem[{\citenamefont{Colangelo et~al.}(2018)\citenamefont{Colangelo, Lanz,
  Leutwyler, and Passemar}}]{Colangelo:2018jxw}
\bibinfo{author}{\bibfnamefont{G.}~\bibnamefont{Colangelo}},
  \bibinfo{author}{\bibfnamefont{S.}~\bibnamefont{Lanz}},
  \bibinfo{author}{\bibfnamefont{H.}~\bibnamefont{Leutwyler}},
  \bibnamefont{and} \bibinfo{author}{\bibfnamefont{E.}~\bibnamefont{Passemar}},
  \bibinfo{journal}{Eur. Phys. J. C} \textbf{\bibinfo{volume}{78}},
  \bibinfo{pages}{947} (\bibinfo{year}{2018}), \eprint{1807.11937}.

\bibitem[{\citenamefont{Seng et~al.}(2021{\natexlab{a}})\citenamefont{Seng,
  Galviz, Gorchtein, and Mei\ss{}ner}}]{Seng:2021boy}
\bibinfo{author}{\bibfnamefont{C.-Y.} \bibnamefont{Seng}},
  \bibinfo{author}{\bibfnamefont{D.}~\bibnamefont{Galviz}},
  \bibinfo{author}{\bibfnamefont{M.}~\bibnamefont{Gorchtein}},
  \bibnamefont{and} \bibinfo{author}{\bibfnamefont{U.~G.}
  \bibnamefont{Mei\ss{}ner}}, \bibinfo{journal}{Phys. Lett. B}
  \textbf{\bibinfo{volume}{820}}, \bibinfo{pages}{136522}
  (\bibinfo{year}{2021}{\natexlab{a}}), \eprint{2103.00975}.

\bibitem[{\citenamefont{Seng et~al.}(2021{\natexlab{b}})\citenamefont{Seng,
  Galviz, Gorchtein, and Mei\ss{}ner}}]{Seng:2021wcf}
\bibinfo{author}{\bibfnamefont{C.-Y.} \bibnamefont{Seng}},
  \bibinfo{author}{\bibfnamefont{D.}~\bibnamefont{Galviz}},
  \bibinfo{author}{\bibfnamefont{M.}~\bibnamefont{Gorchtein}},
  \bibnamefont{and} \bibinfo{author}{\bibfnamefont{U.-G.}
  \bibnamefont{Mei\ss{}ner}}, \bibinfo{journal}{JHEP}
  \textbf{\bibinfo{volume}{11}}, \bibinfo{pages}{172}
  (\bibinfo{year}{2021}{\natexlab{b}}), \eprint{2103.04843}.

\bibitem[{\citenamefont{Seng et~al.}(2020{\natexlab{b}})\citenamefont{Seng,
  Galviz, and Mei\ss{}ner}}]{Seng:2019lxf}
\bibinfo{author}{\bibfnamefont{C.-Y.} \bibnamefont{Seng}},
  \bibinfo{author}{\bibfnamefont{D.}~\bibnamefont{Galviz}}, \bibnamefont{and}
  \bibinfo{author}{\bibfnamefont{U.-G.} \bibnamefont{Mei\ss{}ner}},
  \bibinfo{journal}{JHEP} \textbf{\bibinfo{volume}{02}}, \bibinfo{pages}{069}
  (\bibinfo{year}{2020}{\natexlab{b}}), \eprint{1910.13208}.

\bibitem[{\citenamefont{Seng et~al.}(2020{\natexlab{c}})\citenamefont{Seng,
  Feng, Gorchtein, Jin, and Mei\ss{}ner}}]{Seng:2020jtz}
\bibinfo{author}{\bibfnamefont{C.-Y.} \bibnamefont{Seng}},
  \bibinfo{author}{\bibfnamefont{X.}~\bibnamefont{Feng}},
  \bibinfo{author}{\bibfnamefont{M.}~\bibnamefont{Gorchtein}},
  \bibinfo{author}{\bibfnamefont{L.-C.} \bibnamefont{Jin}}, \bibnamefont{and}
  \bibinfo{author}{\bibfnamefont{U.-G.} \bibnamefont{Mei\ss{}ner}},
  \bibinfo{journal}{JHEP} \textbf{\bibinfo{volume}{10}}, \bibinfo{pages}{179}
  (\bibinfo{year}{2020}{\natexlab{c}}), \eprint{2009.00459}.

\bibitem[{\citenamefont{Sirlin}(1978)}]{Sirlin:1977sv}
\bibinfo{author}{\bibfnamefont{A.}~\bibnamefont{Sirlin}},
  \bibinfo{journal}{Rev. Mod. Phys.} \textbf{\bibinfo{volume}{50}},
  \bibinfo{pages}{573} (\bibinfo{year}{1978}), \bibinfo{note}{[Erratum: Rev.
  Mod. Phys.50,905(1978)]}.

\bibitem[{\citenamefont{Seng}(2021)}]{Seng:2021syx}
\bibinfo{author}{\bibfnamefont{C.-Y.} \bibnamefont{Seng}},
  \bibinfo{journal}{Particles} \textbf{\bibinfo{volume}{4}},
  \bibinfo{pages}{397} (\bibinfo{year}{2021}), \eprint{2108.03279}.

\bibitem[{\citenamefont{Feng et~al.}(2020)\citenamefont{Feng, Gorchtein, Jin,
  Ma, and Seng}}]{Feng:2020zdc}
\bibinfo{author}{\bibfnamefont{X.}~\bibnamefont{Feng}},
  \bibinfo{author}{\bibfnamefont{M.}~\bibnamefont{Gorchtein}},
  \bibinfo{author}{\bibfnamefont{L.-C.} \bibnamefont{Jin}},
  \bibinfo{author}{\bibfnamefont{P.-X.} \bibnamefont{Ma}}, \bibnamefont{and}
  \bibinfo{author}{\bibfnamefont{C.-Y.} \bibnamefont{Seng}},
  \bibinfo{journal}{Phys. Rev. Lett.} \textbf{\bibinfo{volume}{124}},
  \bibinfo{pages}{192002} (\bibinfo{year}{2020}), \eprint{2003.09798}.

\bibitem[{\citenamefont{Ma et~al.}(2021)\citenamefont{Ma, Feng, Gorchtein, Jin,
  and Seng}}]{Ma:2021azh}
\bibinfo{author}{\bibfnamefont{P.-X.} \bibnamefont{Ma}},
  \bibinfo{author}{\bibfnamefont{X.}~\bibnamefont{Feng}},
  \bibinfo{author}{\bibfnamefont{M.}~\bibnamefont{Gorchtein}},
  \bibinfo{author}{\bibfnamefont{L.-C.} \bibnamefont{Jin}}, \bibnamefont{and}
  \bibinfo{author}{\bibfnamefont{C.-Y.} \bibnamefont{Seng}},
  \bibinfo{journal}{Phys. Rev. D} \textbf{\bibinfo{volume}{103}},
  \bibinfo{pages}{114503} (\bibinfo{year}{2021}), \eprint{2102.12048}.

\bibitem[{\citenamefont{Cirigliano et~al.}(2002)\citenamefont{Cirigliano,
  Knecht, Neufeld, Rupertsberger, and Talavera}}]{Cirigliano:2001mk}
\bibinfo{author}{\bibfnamefont{V.}~\bibnamefont{Cirigliano}},
  \bibinfo{author}{\bibfnamefont{M.}~\bibnamefont{Knecht}},
  \bibinfo{author}{\bibfnamefont{H.}~\bibnamefont{Neufeld}},
  \bibinfo{author}{\bibfnamefont{H.}~\bibnamefont{Rupertsberger}},
  \bibnamefont{and} \bibinfo{author}{\bibfnamefont{P.}~\bibnamefont{Talavera}},
  \bibinfo{journal}{Eur. Phys. J.} \textbf{\bibinfo{volume}{C23}},
  \bibinfo{pages}{121} (\bibinfo{year}{2002}), \eprint{hep-ph/0110153}.

\bibitem[{\citenamefont{Cirigliano et~al.}(2004)\citenamefont{Cirigliano,
  Neufeld, and Pichl}}]{Cirigliano:2004pv}
\bibinfo{author}{\bibfnamefont{V.}~\bibnamefont{Cirigliano}},
  \bibinfo{author}{\bibfnamefont{H.}~\bibnamefont{Neufeld}}, \bibnamefont{and}
  \bibinfo{author}{\bibfnamefont{H.}~\bibnamefont{Pichl}},
  \bibinfo{journal}{Eur. Phys. J. C} \textbf{\bibinfo{volume}{35}},
  \bibinfo{pages}{53} (\bibinfo{year}{2004}), \eprint{hep-ph/0401173}.

\bibitem[{\citenamefont{Cirigliano et~al.}(2008)\citenamefont{Cirigliano,
  Giannotti, and Neufeld}}]{Cirigliano:2008wn}
\bibinfo{author}{\bibfnamefont{V.}~\bibnamefont{Cirigliano}},
  \bibinfo{author}{\bibfnamefont{M.}~\bibnamefont{Giannotti}},
  \bibnamefont{and} \bibinfo{author}{\bibfnamefont{H.}~\bibnamefont{Neufeld}},
  \bibinfo{journal}{JHEP} \textbf{\bibinfo{volume}{11}}, \bibinfo{pages}{006}
  (\bibinfo{year}{2008}), \eprint{0807.4507}.

\bibitem[{\citenamefont{Bloch and Nordsieck}(1937)}]{Bloch:1937pw}
\bibinfo{author}{\bibfnamefont{F.}~\bibnamefont{Bloch}} \bibnamefont{and}
  \bibinfo{author}{\bibfnamefont{A.}~\bibnamefont{Nordsieck}},
  \bibinfo{journal}{Phys. Rev.} \textbf{\bibinfo{volume}{52}},
  \bibinfo{pages}{54} (\bibinfo{year}{1937}).

\bibitem[{\citenamefont{Yennie et~al.}(1961)\citenamefont{Yennie, Frautschi,
  and Suura}}]{Yennie:1961ad}
\bibinfo{author}{\bibfnamefont{D.~R.} \bibnamefont{Yennie}},
  \bibinfo{author}{\bibfnamefont{S.~C.} \bibnamefont{Frautschi}},
  \bibnamefont{and} \bibinfo{author}{\bibfnamefont{H.}~\bibnamefont{Suura}},
  \bibinfo{journal}{Annals Phys.} \textbf{\bibinfo{volume}{13}},
  \bibinfo{pages}{379} (\bibinfo{year}{1961}).

\bibitem[{\citenamefont{Kinoshita and Sirlin}(1959)}]{Kinoshita:1958ru}
\bibinfo{author}{\bibfnamefont{T.}~\bibnamefont{Kinoshita}} \bibnamefont{and}
  \bibinfo{author}{\bibfnamefont{A.}~\bibnamefont{Sirlin}},
  \bibinfo{journal}{Phys. Rev.} \textbf{\bibinfo{volume}{113}},
  \bibinfo{pages}{1652} (\bibinfo{year}{1959}).

\bibitem[{\citenamefont{Kinoshita}(1962)}]{Kinoshita:1962ur}
\bibinfo{author}{\bibfnamefont{T.}~\bibnamefont{Kinoshita}},
  \bibinfo{journal}{J. Math. Phys.} \textbf{\bibinfo{volume}{3}},
  \bibinfo{pages}{650} (\bibinfo{year}{1962}).

\bibitem[{\citenamefont{Lee and Nauenberg}(1964)}]{Lee:1964is}
\bibinfo{author}{\bibfnamefont{T.~D.} \bibnamefont{Lee}} \bibnamefont{and}
  \bibinfo{author}{\bibfnamefont{M.}~\bibnamefont{Nauenberg}},
  \bibinfo{journal}{Phys. Rev.} \textbf{\bibinfo{volume}{133}},
  \bibinfo{pages}{B1549} (\bibinfo{year}{1964}).

\bibitem[{\citenamefont{Roos and Sirlin}(1971)}]{Roos:1971mj}
\bibinfo{author}{\bibfnamefont{M.}~\bibnamefont{Roos}} \bibnamefont{and}
  \bibinfo{author}{\bibfnamefont{A.}~\bibnamefont{Sirlin}},
  \bibinfo{journal}{Nucl. Phys. B} \textbf{\bibinfo{volume}{29}},
  \bibinfo{pages}{296} (\bibinfo{year}{1971}).

\bibitem[{\citenamefont{Marciano}(1975)}]{Marciano:1975de}
\bibinfo{author}{\bibfnamefont{W.~J.} \bibnamefont{Marciano}},
  \bibinfo{journal}{Phys. Rev. D} \textbf{\bibinfo{volume}{12}},
  \bibinfo{pages}{3861} (\bibinfo{year}{1975}).

\bibitem[{\citenamefont{Meister and Yennie}(1963)}]{Meister:1963zz}
\bibinfo{author}{\bibfnamefont{N.}~\bibnamefont{Meister}} \bibnamefont{and}
  \bibinfo{author}{\bibfnamefont{D.}~\bibnamefont{Yennie}},
  \bibinfo{journal}{Phys. Rev.} \textbf{\bibinfo{volume}{130}},
  \bibinfo{pages}{1210} (\bibinfo{year}{1963}).

\bibitem[{\citenamefont{Bytev et~al.}(2003)\citenamefont{Bytev, Kuraev, Baratt,
  and Thompson}}]{Bytev:2002nx}
\bibinfo{author}{\bibfnamefont{V.}~\bibnamefont{Bytev}},
  \bibinfo{author}{\bibfnamefont{E.}~\bibnamefont{Kuraev}},
  \bibinfo{author}{\bibfnamefont{A.}~\bibnamefont{Baratt}}, \bibnamefont{and}
  \bibinfo{author}{\bibfnamefont{J.}~\bibnamefont{Thompson}},
  \bibinfo{journal}{Eur. Phys. J. C} \textbf{\bibinfo{volume}{27}},
  \bibinfo{pages}{57} (\bibinfo{year}{2003}), \bibinfo{note}{[Erratum:
  Eur.Phys.J.C 34, 523--524 (2004)]}, \eprint{hep-ph/0210049}.

\bibitem[{\citenamefont{'t~Hooft and Veltman}(1972)}]{tHooft:1972tcz}
\bibinfo{author}{\bibfnamefont{G.}~\bibnamefont{'t~Hooft}} \bibnamefont{and}
  \bibinfo{author}{\bibfnamefont{M.~J.~G.} \bibnamefont{Veltman}},
  \bibinfo{journal}{Nucl. Phys. B} \textbf{\bibinfo{volume}{44}},
  \bibinfo{pages}{189} (\bibinfo{year}{1972}).

\bibitem[{\citenamefont{Bollini and Giambiagi}(1972)}]{Bollini:1972ui}
\bibinfo{author}{\bibfnamefont{C.~G.} \bibnamefont{Bollini}} \bibnamefont{and}
  \bibinfo{author}{\bibfnamefont{J.~J.} \bibnamefont{Giambiagi}},
  \bibinfo{journal}{Nuovo Cim. B} \textbf{\bibinfo{volume}{12}},
  \bibinfo{pages}{20} (\bibinfo{year}{1972}).

\bibitem[{\citenamefont{Ashmore}(1972)}]{Ashmore:1972uj}
\bibinfo{author}{\bibfnamefont{J.~F.} \bibnamefont{Ashmore}},
  \bibinfo{journal}{Lett. Nuovo Cim.} \textbf{\bibinfo{volume}{4}},
  \bibinfo{pages}{289} (\bibinfo{year}{1972}).

\bibitem[{\citenamefont{Gastmans and Meuldermans}(1973)}]{Gastmans:1973uv}
\bibinfo{author}{\bibfnamefont{R.}~\bibnamefont{Gastmans}} \bibnamefont{and}
  \bibinfo{author}{\bibfnamefont{R.}~\bibnamefont{Meuldermans}},
  \bibinfo{journal}{Nucl. Phys. B} \textbf{\bibinfo{volume}{63}},
  \bibinfo{pages}{277} (\bibinfo{year}{1973}).

\bibitem[{\citenamefont{Marciano and Sirlin}(1975)}]{Marciano:1974tv}
\bibinfo{author}{\bibfnamefont{W.}~\bibnamefont{Marciano}} \bibnamefont{and}
  \bibinfo{author}{\bibfnamefont{A.}~\bibnamefont{Sirlin}},
  \bibinfo{journal}{Nucl. Phys. B} \textbf{\bibinfo{volume}{88}},
  \bibinfo{pages}{86} (\bibinfo{year}{1975}).

\bibitem[{\citenamefont{Marques and Papanicolaou}(1975)}]{Marques:1974ab}
\bibinfo{author}{\bibfnamefont{G.~C.} \bibnamefont{Marques}} \bibnamefont{and}
  \bibinfo{author}{\bibfnamefont{N.}~\bibnamefont{Papanicolaou}},
  \bibinfo{journal}{Phys. Rev. D} \textbf{\bibinfo{volume}{12}},
  \bibinfo{pages}{1052} (\bibinfo{year}{1975}).

\bibitem[{\citenamefont{Marciano et~al.}(1975)\citenamefont{Marciano, Marques,
  and Papanicolaou}}]{Marciano:1974kw}
\bibinfo{author}{\bibfnamefont{W.~J.} \bibnamefont{Marciano}},
  \bibinfo{author}{\bibfnamefont{G.~C.} \bibnamefont{Marques}},
  \bibnamefont{and}
  \bibinfo{author}{\bibfnamefont{N.}~\bibnamefont{Papanicolaou}},
  \bibinfo{journal}{Nucl. Phys. B} \textbf{\bibinfo{volume}{96}},
  \bibinfo{pages}{237} (\bibinfo{year}{1975}).

\bibitem[{\citenamefont{Collins}(2013)}]{Collins:2011zzd}
\bibinfo{author}{\bibfnamefont{J.}~\bibnamefont{Collins}},
  \emph{\bibinfo{title}{{Foundations of perturbative QCD}}},
  vol.~\bibinfo{volume}{32} (\bibinfo{publisher}{Cambridge University Press},
  \bibinfo{year}{2013}), ISBN \bibinfo{isbn}{978-1-107-64525-7,
  978-1-107-64525-7, 978-0-521-85533-4, 978-1-139-09782-6}.

\bibitem[{\citenamefont{Becher et~al.}(2015)\citenamefont{Becher, Broggio, and
  Ferroglia}}]{Becher:2014oda}
\bibinfo{author}{\bibfnamefont{T.}~\bibnamefont{Becher}},
  \bibinfo{author}{\bibfnamefont{A.}~\bibnamefont{Broggio}}, \bibnamefont{and}
  \bibinfo{author}{\bibfnamefont{A.}~\bibnamefont{Ferroglia}},
  \emph{\bibinfo{title}{{Introduction to Soft-Collinear Effective Theory}}},
  vol. \bibinfo{volume}{896} (\bibinfo{publisher}{Springer},
  \bibinfo{year}{2015}), \eprint{1410.1892}.

\bibitem[{\citenamefont{Gorsky et~al.}(1989)\citenamefont{Gorsky, Ioffe, and
  Khodjamirian}}]{Gorsky:1989qd}
\bibinfo{author}{\bibfnamefont{A.~S.} \bibnamefont{Gorsky}},
  \bibinfo{author}{\bibfnamefont{B.~L.} \bibnamefont{Ioffe}}, \bibnamefont{and}
  \bibinfo{author}{\bibfnamefont{A.~Y.} \bibnamefont{Khodjamirian}},
  \bibinfo{journal}{Phys. Lett. B} \textbf{\bibinfo{volume}{227}},
  \bibinfo{pages}{474} (\bibinfo{year}{1989}).

\bibitem[{\citenamefont{Smilga}(1991)}]{Smilga:1990uq}
\bibinfo{author}{\bibfnamefont{A.~V.} \bibnamefont{Smilga}},
  \bibinfo{journal}{Comments Nucl. Part. Phys.} \textbf{\bibinfo{volume}{20}},
  \bibinfo{pages}{69} (\bibinfo{year}{1991}).

\bibitem[{\citenamefont{Batley et~al.}(2018)}]{NA482:2018rgv}
\bibinfo{author}{\bibfnamefont{J.~R.} \bibnamefont{Batley}}
  \bibnamefont{et~al.} (\bibinfo{collaboration}{NA48/2}),
  \bibinfo{journal}{JHEP} \textbf{\bibinfo{volume}{10}}, \bibinfo{pages}{150}
  (\bibinfo{year}{2018}), \eprint{1808.09041}.

\bibitem[{\citenamefont{Siegel}(1979)}]{Siegel:1979wq}
\bibinfo{author}{\bibfnamefont{W.}~\bibnamefont{Siegel}},
  \bibinfo{journal}{Phys. Lett. B} \textbf{\bibinfo{volume}{84}},
  \bibinfo{pages}{193} (\bibinfo{year}{1979}).

\bibitem[{\citenamefont{Siegel}(1980)}]{Siegel:1980qs}
\bibinfo{author}{\bibfnamefont{W.}~\bibnamefont{Siegel}},
  \bibinfo{journal}{Phys. Lett. B} \textbf{\bibinfo{volume}{94}},
  \bibinfo{pages}{37} (\bibinfo{year}{1980}).

\end{thebibliography}

\end{document}